\documentclass[twocolumn,twoside,9pt]{IEEEtran}
\usepackage{graphicx,epstopdf,multirow,booktabs,pgfplots}

\usepackage{amsmath,bm}
\usepackage{amssymb,amsthm}
\usepackage[T1]{fontenc}
\usepackage[backref=page,colorlinks=true,citecolor=cyan,linkcolor=magenta]{hyperref}
\allowdisplaybreaks
\usepackage{setspace}
\usepackage{pgfplotstable}
\usepackage{pgfplots}
\usepackage{enumitem}
\usepackage{cite}
\pgfplotsset{compat=1.13}
\usepackage{multirow,multicol,subfigure,booktabs}

\usepackage{amssymb}
\usepackage{amsmath}
\usepackage{url}
\usepackage{algorithm,algpseudocode}
\makeatletter
\newcommand\subparagraph{%
  \@startsection{subparagraph}{5}
  {\parindent}
  {3.25ex \@plus 1ex \@minus .2ex}
  {-1em}
  {\normalfont\normalsize\bfseries}}
\makeatother
\usepackage[compact]{titlesec}
\let\subparagraph\relax % You don't want to use \subparagraph

\titlespacing{\section}{0pt}{2ex}{1ex}
\titlespacing{\subsection}{0pt}{1ex}{1ex}
\titlespacing{\subsubsection}{0pt}{0.5ex}{0ex}

\usepackage{mathptmx}
\usepackage{color}

\newcommand{\kaarthik}[1]{\textcolor{black}{#1}}
\newcommand{\linecomment}[1]{\textcolor{black}{#1}}

\usepackage{accents}

\usepackage{hyperref}
\allowdisplaybreaks

\theoremstyle{remark}
\newtheorem{proposition}{Proposition}

\usepackage{filecontents}
\usepackage{etoolbox}
\usepackage{tabularx}

\graphicspath{figures/}
\begin{document}
% \doublespacing
\title{\Large \bf Chance-Constrained Unit Commitment with N-1 Security and Wind Uncertainty}

\author{Kaarthik Sundar$^\dag$, Harsha Nagarajan$^\dag$, Line Roald$^\dag$, Sidhant Misra$^\dag$, Russell Bent$^\dag$, Daniel Bienstock$^\ddag$
\thanks{$\dag$ Center for Nonlinear Studies, Los Alamos National Laboratory, Los Alamos, NM. Contact: \texttt{rbent@lanl.gov} \newline
$\ddag$ Department of Industrial Engineering and Operations Research and Department of Applied Physics and Applied Mathematics, Columbia University, NY.}\;
}
\makeatletter
\patchcmd{\@maketitle}
  {\addvspace{0.5\baselineskip}\egroup}
  {\addvspace{-1\baselineskip}\egroup}
  {}
  {}
\makeatother

\markboth{Journal of \LaTeX\ Class Files,~Vol.~14, No.~8, November~2016}%
{Sundar \MakeLowercase{\textit{et al.}}: Chance-Constrained Unit Commitment with N-1 Security and Wind Uncertainty}

\maketitle

\begin{abstract}
As renewable wind energy penetration rates continue to increase, one of the major challenges facing grid operators is the question of how to control transmission grids in a reliable and a cost-efficient manner. The stochastic nature of wind forces an alteration of traditional methods for solving day-ahead and look-ahead unit commitment and dispatch. In particular, the variability of wind generation increases the risk of unexpected overloads and cascading events. To address these questions, we present an N-1 Security and Chance-Constrained Unit Commitment (SCCUC) that includes models of generation reserves that respond to wind fluctuations and component outages. We formulate the SCCUC as a mixed-integer, second-order cone problem that limits the probability of failure. We develop a modified Benders decomposition algorithm to solve the problem to optimality and present detailed case studies on the IEEE RTS-96 three-area and the IEEE 300 NESTA test systems. The case studies assess the economic impacts of contingencies and various degrees of wind power penetration and demonstrate the effectiveness and scalability of the algorithm.
\end{abstract}

\IEEEpeerreviewmaketitle

\section{Introduction}
\label{sec:intro}

Transmission grids play a vital role in the reliable and economically efficient operation of electric power systems. As renewable energy penetration rates continue to grow, the stochastic nature of renewable generation necessitates an alteration of traditional methods for solving day-ahead Unit Commitment (UC) and generation dispatch. The fluctuations caused by wind and solar generation can bring system components closer to their physical limits or cause unforeseen overloads in real-time operation. This reduces the ability to adequately handle contingencies and increases operational risk. As a result, power system operators are interested in securing the grid against component failures in the presence of these fluctuations. 

More formally, the security of a power grid is defined by its ability to survive contingencies \cite{Kundur2004}.
The failure to secure a power system could potentially result in cascading events and large scale blackouts \cite{Andersson2005,CIGRE2009}. The N-1 security criterion was developed to avoid such incidents (see \cite{Stott1987} and references therein). 
While N-1 is an important security criteria, it is only well defined for a deterministic operating condition. 
With uncertainty from wind and solar power, the system might face considerable risk from cascading events in real-time operation, 
even though the system was N-1 secure for the forecast operating condition. Thus, there is a need for methods that appropriately limit the risk from uncertain deviations. 

The literature considers a number of approaches to handle uncertainty in power systems operational planning, including 
two-stage and multi-stage stochastic programming approaches \cite{Bouffard2008,Morales2009,Ucckun2016}, robust models \cite{Lorca2015,Warrington2013} and chance-constrained formulations \cite{Margellos2013,Vrakopoulou2013probabilisticreserve,Roald2013,Bienstock2014,Sundar2016}.
Here, we use the latter approach, as chance constraints allows for a comprehensive, yet computationally tractable uncertainty representation and aligns well with several methods applied in industry. Industry practices that are naturally represented through chance constraints include probabilistic reserve dimensioning (applied in ENTSO-E \cite{ENTSOE2013supportingLFCR}) and the definitions of reliability margins for flow-based market coupling in parts of Europe \cite{CWE2011}.
We develop a comprehensive model that incorporates all aspects of day-ahead planning and security discussed above - UC for generators, N-1 security constraints for line and generator outages, chance constraints to ensure stochastic security with respect to wind, as well as constraints that model \linecomment{the activation of generation reserves for both wind fluctuations and outages}. We refer to this problem as the N-1 Security and Chance-Constrained Unit Commitment (SCCUC) problem. 

In this paper, we utilize the model that incorporates all the aforementioned aspects and addresses the computational issues associated with scalability with a new, modified Benders decomposition algorithm. From a scaling perspective, other stochastic UC models can handle problems with 225 \cite{Papavasiliou2013} and 118 buses \cite{Wang2012}. While these models are comparable in size to our own, they do not include N-1 security, which increases the size of the problem by an order of magnitude.
Thus, we argue that our model and approach is more scalable than these prior approaches.

We now review the models and algorithms that have addressed variations of unit commitment and optimal power flow problems that are most similar to the SCCUC. In \cite{Papavasiliou2013}, a stochastic UC variant with generator and line contingencies was considered; the authors develop a two-stage, dual-decomposition algorithm that includes wind with a scenario-based approach. The approach was tested on a 225-bus model of the California power system and the average computation time for solving a 42-scenario model was approximately 6 hours. A variant of the problem in \cite{Papavasiliou2013} was studied both with and without generation reserve modeling in \cite{Wang2012} and \cite{Margellos2013}, respectively. The authors in \cite{Margellos2013} also developed a sampling-based approach to account for wind uncertainty with a chance-constrained formulation. They compared their algorithms, using Monte Carlo simulations, with a deterministic variant on a modified IEEE 118-bus network and the IEEE 30-bus network, respectively. Authors in \cite{Pandzic2015} developed a transmission-constrained UC formulation where the wind uncertainty is modeled using an interval formulation. They compared the solution cost and robustness of their approach with existing stochastic, interval and robust UC techniques  on the IEEE RTS-96 test system. Finally, \cite{Zhao2013} developed a multi-stage robust UC formulation that considers uncertainties due to both demand and wind and tested the effectiveness of the algorithm on the IEEE 118-bus system. 

Within the literature, the number of papers that consider  stochastic versions of UC with N-1 contingency modeling is limited. For instance, the chance-constrained OPF problem of \cite{Vrakopoulou2013} incorporates both wind and N-1 security constraints, but omits UC modeling. The lack of discrete variables associated with UC models greatly reduces the problem complexity. \cite{Nasri2014} develops a Benders decomposition algorithm to solve the UC with wind uncertainty and AC power-flow constraints, but omits the N-1 contingency modeling. \cite{Vrakopoulou2013} and \cite{Nasri2014} study the effectiveness of their proposed approaches on the IEEE 30-bus and IEEE 96-RTS one area system, respectively.  \cite{Moreira2015} develops a Benders decomposition algorithm for the UC with N-1 contingency modeling, reserve modeling, uncertainty in demands and illustrate their algorithm on a 3-bus and the IEEE 24-bus systems. \cite{Pozo2013} develops a chance-constrained model with $N$-$k$ security criterion; they handle the chance constraints by drawing $100$ samples from the probability distribution for wind and enforce the hard constraints on all the $100$ scenarios. Their algorithms are tested on a $3$-bus and a $10$-bus model. In terms of model comprehensiveness, the models most similar to ours is that of \cite{Bertsimas2013} and \cite{Sundar2016}. The authors in \cite{Bertsimas2013} developed a two-stage, adaptive robust UC model with security constraints and nodal net injection uncertainty. The model includes a deterministic uncertainty set, unlike the probability distribution of this paper, to model the wind. They proposed a Benders-based decomposition algorithm to handle line contingencies over a large-scale system, however they do not consider generator contingencies. It is the combination of both types of contingencies, especially the generator contingencies with reserve modeling, that makes our problem closer to the necessities of an Independent System Operator (ISO) and a lot more difficult to solve.

In this paper, we formulate SCCUC as a large Mixed-Integer Second-Order Cone Program (MISOCP) that leverages the recent work in \cite{Bienstock2014, Lubin2015, Roald2013, Sundar2016}. This formulation \linecomment{considers two kinds of reserves, which respond to wind fluctuations and outages, respectively}. We develop a sequential, modified Benders decomposition algorithm that exploits the block diagonal structure of the constraint matrix. The decomposition algorithm solves an inner problem that accounts for generator contingencies using a Benders decomposition. The outer loop accounts for line contingency constraints which are modeled as Second-Order Cones (SOCs) and efficiently solved via an outer-approximation technique. 

Our algorithm is parallelizable and extensive computational experiments based on the three area IEEE RTS-96 system and IEEE 300 NESTA test system are used to demonstrate the effectiveness of the algorithm. We also present a detailed comparison of the SCCUC formulation to the deterministic variant of the problem to illustrate the economic and operational advantages of the SCCUC.

\section{Problem Formulation} \label{sec:formulation}

Throughout this paper we utilize the linearized DC power flow model. While this model has many limitations, it is the model most generally used in the UC literature. In order to isolate the primary complexities (chance-constrained response of wind fluctuations and the combinatorial explosion of solutions) and highlight this paper's core algorithmic contributions, we adhere to the traditional UC model with linearized DC power flow equations. Future work will need to consider more realistic power flow models and develop tractable chance-constrained formulations on models of the nonlinear power flow physics \cite{lubin2015twoside}.  

\subsection{Nomenclature}
\label{sec:nomenclature}
{\small
\noindent \textit{\underline{\smash{Sets}}}\\
$\mathcal B$ - set of buses, indexed by $b$ \\
$\mathcal L$ - set of lines, indexed by $\ell$ \\
$\mathcal G$ - set of generators, indexed by $i$ \\
$\mathcal S_i$ - set of start-up cost blocks for generator $i$, indexed by $s$ \\
$\mathcal K$ - set of generator cost blocks for generator $i$, indexed by $k$\\
$\mathcal W$ - set of buses with wind generation \\
$\mathcal T$ - set of discretized times (hours), indexed by $t$ \\
$\mathcal C_g$ - set of generator contingencies, indexed by $gc$ \\
$\mathcal C_{\ell}$ - set of line contingencies, indexed by $lc$ \\
\noindent \textit{\underline{\smash{Binary decision variables}}} \\
$x_i(t)$ - generator on-off status at time $t$ \\
$y_i(t)$ - generator 0-1 start-up status at time $t$\\
$z_i(t)$ - generator 0-1 shut-down status at time $t$\\
$w_{is}(t)$ - generator 0-1 status if $i$ starts at time $t$ after being down for $s$ hours\\%start-up block identification; $1$ if $i$ is started up at the beginning of hour $t$ after being down for $s$ hours, $0$ otherwise\\
\noindent \textit{\underline{\smash{Continuous decision variables}}} \\
$sc_i(t)$ - start-up cost of generator $i$ at time $t$, \$\\
$p_i(t)$ - power output of generator $i$ at time $t$, MW \\
$r^+_i(t)$, $r^-_i(t)$ - up/down-reserve of generator $i$ at time $t$, MW\\
% $r^-_i(t)$ - down reserve power output of $i$ at time $t$, MW\\
$r^{out}_i(t)$ - contingency reserve at generator $i$ at time $t$, MW\\
$g^k_i(t)$ - power output on segment $k$ of cost curve of generator $i$ at time $t$, MW\\
$\alpha_i(t)$ - participation factor of generator $i$ at time $t$ \\
$f_{\ell}(t)$ - real power flow of $\ell$ at time $t$, MW \\
$\delta_i^{gc}(t)$ - power provided by $i$ for contingency $gc$ at time $t$, MW\\
%$\alpha^c_i(t)$ - participation factor of $i$ for contingency $c$ at time $t$ \\
$f_{\ell}^{gc}(t)$ - power flow on $\ell$ for contingency $gc$ at time $t$, MW \\
$f_{\ell}^{\ell c}(t)$ - power flow on $\ell$ for contingency $\ell c$ at time $t$, MW \\
$\theta_b(t)$ - phase angle at bus $b$ at time $t$,\\
\noindent \textit{\underline{\smash{Parameters}}}: \\
$\beta_{\ell}$ - susceptance of line $\ell$ \\
$p_i^{\min}$, $p_i^{\max}$ - min. and max. output of generator $i$, MW \\
% $p_i^{\max}$ - maximum output of generator $i$, MW \\
$p_{i,k}^{\max}$ - max. output of generator $i$ in production cost block $k$, MW\\ 
$d_b(t)$ - demand at bus $b$ for hour $t$, MW \\
$a^0_i$ - no-load cost of the generator $i$, \$ \\
$a^1_i$ - linear cost coefficient for $r_i^+$ and $r_i^-$ for generator $i$, \$/MW \\
$a^2_i$ - linear cost coefficient for $r_i^{out}$ for generator $i$, \$/MW \\
$K_i^k$ - slope of the $k^{th}$ segment of the cost curve for $i$, \$/MW \\
$f_{\ell}^{\max}$ - capacity of line $\ell$, MW \\
$\overline L_i$ -  \kaarthik{time generator $i$ has to run starting from $t=0$, hrs}\\
$\underline L_i$ - \kaarthik{time generator $i$ has to be off starting from $t=0$, hrs}\\
$UT_i$, $DT_i$ - min. up-time and down-time of generator $i$, hrs \\
% $DT_i$ - minimum down-time of $i$, hrs \\
$p^{\operatorname{up,init}}_i$, $p^{\operatorname{down,init}}_i$ - number of time periods generator $i$ has been on and off before $t=0$, hrs \\
$p^{\operatorname{on-off}}_i$ - on-off status of generator $i$ at $t=0$ \\
$p_i(0)$ - power output of generator $i$ at $t=0$, MW\\
$RU_i$, $RD_i$  - ramp-up/down limit of generator $i$, MW/hr\\
$c_{is}$ - cost of block $s$ of stepwise start-up cost function of $i$, \$\\
$\overline T_{is}$, $\underline T_{is}$ - upper and lower limit of block $s$ of the start-up cost of generator $i$, hrs\\
$\mu_b(t)$ - output of wind farm at $b$ for time $t$, MW\\
$\omega_b(t)$ - actual wind deviations from forecast $\mu_b(t)$, at time $t$ \\
$r$ - index of the reference bus \\
$R_i$ - bounds on the reserves that purchased from generator $i$, MW\\
$B$ - bus admittance matrix \\
}

In the rest of this paper, bold symbols denote random variables. In particular, $\boldsymbol \omega_b(t)$ is the random variable that models $\omega_b(t)$ for hour $t$. In the SCCUC, we assume that the deviations, $\boldsymbol\omega_b(t)$ are independent and normally distributed with zero mean and variance $\sigma_b(t)^2$ \cite{Bienstock2014}. The model can include correlations across space, geographically, in $\boldsymbol \omega_b(t)$ similar to the models in \cite{Lubin2015, Roald2016}. These wind deviations drive the random fluctuations of the controllable injections  $\boldsymbol p_i(t)$, and line flows $\boldsymbol{f}_{\ell}(t)$. We remark that, though we assume the wind power generation to have a normal distribution \linecomment{in our chance constraint reformulation, chance constraint reformulations based on first and second moment information are available for more general distributions \cite{Lubin2015, summers2014, Roald2015} and would still give rise to constraints of the same mathematical form.} 

Finally, we let $\boldsymbol \Omega(t) = \sum_{b\in \mathcal W} \boldsymbol \omega(t)$ denote the total deviation in the wind from the forecast. For notational convenience, \linecomment{we use bars to denote \emph{vectors} of variables, i.e., for the power generation $\bar{p}(t)$, forecast wind power $\bar{\mu}(t)$, power demand at the buses $\bar d(t)$, wind deviations $\bar{\omega}(t)$, generation reserve activation during generator contingencies $\bar{\delta}^{gc}(t)$, participation factors of the controllable generators in balancing wind power generation $\bar\alpha(t)$, and generation reserves procured to respond to up and down wind fluctuations $\bar r^+(t)$, $\bar r^-(t)$ and to generator outages $\bar r^{out}(t)$, respectively.}
\linecomment{
\subsection{Generation reserve activation\label{sec:activation}}
In the SCCUC, we model two types of reserves:
\begin{enumerate}
    \item Reserves that continuously maintain balance during wind power fluctuations. These reserves are activated based on the generator participation factor $\bar \alpha_i(t)$ for each generator $i$, as described in Sec. \ref{sec:gencontrol}. The corresponding reserve capacities are denoted by $r^+_i(t), r^-_i(t)$ for up and down capacity, respectively.
    \item Reserves that are activated to restore balance after generator outages. Since generator outages are a finite number of events, we introduce variables $\delta_i^{gc}$ to denote the change in generation at each generator $i$ following the outage $gc$, as described in Sec. \ref{sec:reserves}, with corresponding reserve capacities $r_i^{out}$. 
\end{enumerate}
Note that the two types of reserves are distinct in our formulation, i.e., we are not able to substitute $r_i^{out}$ by $r_i^+$ or vice versa. The linear cost coefficients for purchasing generation reserves from a generator $i$ at time $t$ are given by $a_i^1$ for $r^+_i(t), r^-_i(t)$ and $a_i^2$ for $r_i^{out}$, respectively.
}

\subsubsection{Generation response to wind fluctuations \label{sec:gencontrol}}
Since secure operation of the power system requires balance between produced and consumed power at all times, any deviation \linecomment{of the wind power generation away from the forecast} must be balanced by an adjustment in the controllable generation. We model the adjustments \linecomment{to the wind power fluctuations} as an affine \linecomment{control} policy, reflecting \linecomment{the activation of control reserves through the} automatic generation control (AGC), which establishes power balance within tens
of seconds \cite{Wood2012}:
\begin{flalign}
\boldsymbol p_i(t) = p_i(t) - \alpha_i(t) \boldsymbol \Omega(t). \label{eq:genresponse}
\end{flalign}
Here, $\alpha_i(t) \geqslant 0$ is the participation factor for the controllable generator $i$. When $\sum_i \alpha_i(t) = 1$, Eq. \eqref{eq:genresponse} guarantees balance of generation and load for every time period $t$. \kaarthik{In many electric utilities, $\bar{\alpha}(t)$ does not vary with time, which is a special case of this more general model.} We remark that computational tractability of the SCCUC is improved considerably when $\bar{\alpha}(t)$ is a constant or \linecomment{is the same over the time horizon $t\in\mathcal T$}. However, \linecomment{it has been shown to be economically advantageous to optimize the reserve activation by including $\bar{\alpha}(t)$ as an optimization variable} (see \cite{Bienstock2014, Roald2016EnergyCon, Roald2016}) and hence, we let $\alpha_i(t)$ be a variable for every generator $i\in \mathcal G$ and time period $t$.  

\subsubsection{Generation response to generator outages\label{sec:reserves}}
Generator $i$'s output after the outage of a generator $gc$ during hour $t$ is modeled as
\begin{flalign}
p^{gc}_i(t) = p_i(t) + \delta^{gc}_i(t) \label{eq:genresponsecontingency}
\end{flalign}
To ensure power balance we enforce 
\begin{flalign}
\sum_{i \in \mathcal G} \delta^{gc}_i(t) = 0 \text{ and } \delta^{gc}_{gc}(t) = - p_{gc}(t),  \label{eq:powerbalancegout}
\end{flalign}
\linecomment{where the latter constraints enforce the change in the power output of the outaged generator to be equal to its pre-contingency generation.} 

We remark that our formulations ignore the impact of wind fluctuations in the case of generator contingencies. \linecomment{Specifically, we neglect the impact of losing the reserves that were activated to deal with wind power fluctuations at the outaged generator (i.e., after the outage of a generator that participated in wind power balancing, the participation factors of the remaining online generators do not sum to one, $\sum_{i\in\mathcal{G}\backslash gc} \alpha_i(t) \neq 1$). This leaves an increased risk of not having sufficient reserves available to cover wind power fluctuations. Furthermore, there is a chance of post-contingency line overloading due to combinations of generation outages and wind power fluctuations that have not been accounted for in the model. In situations where the wind power fluctuations are not particularly large, we assume that the impact on the system is benign, as the other generators can increase their reserve activation to balance wind fluctuations and the power flows will not deviate too far from the scheduled flows.} We assume that the probability of a large generation outage happening simultaneously with a large wind deviation is comparatively rare, and hence outside the scope of normal N-1 security assessment. 
\linecomment{
\subsubsection{Generation response to line outages \label{sec:genlinecontingencies}}
In the case of a line contingency, there is no immediate load-generation imbalance in the system (unless the outage is a radial line disconnecting a generator, in which case it is treated as a generator outage). Since our reserve modelling in this paper is focused mainly on maintaining power balance in the system, we assume that the generators continue to generate power according to their pre-contingency generation levels. 
Conceptually, it is easy to extend the modelling to include post-contingency generation redispatch in a similar manner as for the generator outage response described in \eqref{eq:genresponsecontingency}, \eqref{eq:powerbalancegout}, i.e., by introducing new variables $\delta_i^{lc}(t)$ to represent the change in generation as in \cite{Moreira2015}. Post-contingency generation redispatch is expected to lower the cost of operation, as it adds flexibility to the system. However, while it is conceptually easy to include these variables, it is computationally challenging due to the large number of additional optimization variables. Further, many regulators require a distinction between reserves (used for balancing) and redispatch (used for congestion management). We therefore choose to focus on the former in this paper, and delegate the latter for future work. 
}
\subsection{Line flows \label{sec:lineflows}}
\subsubsection{Normal operation \label{sec:normal}}
\linecomment{With changes in the wind power generation and subsequent reserve activation according to \eqref{eq:genresponse}, the power flows across the lines also vary.} These random fluctuations in line flows, $\boldsymbol f_{\ell}(t)$ for line $\ell$, depend on the wind fluctuations implicitly through the random bus phase angles $\boldsymbol \theta_{b}(t)$, i.e. \cite{Wood2012},
\begin{equation}
\bar p(t) + \bar{\mu}(t) - \bar d(t) + \bar{\boldsymbol \omega}(t) - \boldsymbol \Omega(t)\bar\alpha(t) = B\boldsymbol\theta(t).
\end{equation}
The bus admittance matrix $B$ is invertible after removing the row and column corresponding to the reference bus $r \in \mathcal{B}$, \linecomment{which allows us to express the phase angles $\theta$ as linear functions of the power injections}. The flows $\boldsymbol f_{\ell}(t)$ are linear functions of the bus phase angles, \linecomment{and by substituting the relationship for the angles as a function of the power injections, we obtain} the $|\mathcal L| \times |\mathcal B|$ matrix $M$ as the linear map from power injections to line flows.
Then, the random line flows for each line $\ell$ are computed as:
\begin{flalign}
\boldsymbol f_{\ell}(t) = M_{(\ell,\cdot)} \left(\bar p(t) + \bar{\mu}(t) - \bar d(t) + \bar{\boldsymbol \omega}(t) - \boldsymbol \Omega(t) \bar\alpha(t)\right). \label{eq:randomlineflow}
\end{flalign}  

\subsubsection{Post generator contingencies}\label{sec:gout}
\linecomment{Since we ignore the impact of wind power fluctuations on system after a generator outage, see Eq. \eqref{eq:genresponsecontingency},} the post generator contingency line flows are deterministic, i.e. 
\begin{flalign}
f^{gc}_{\ell}(t) = M_{(\ell,\cdot)} \left(\bar p(t) + \bar \delta^{gc}(t) + \bar{\mu}(t) - \bar d(t) \right) \label{eq:detlineflowcontingency}
\end{flalign}
where $M$ is the matrix defined in Sec. \ref{sec:lineflows} and $f_{\ell}^{gc}(t)$ is the line flow on the line $\ell$ post generator outage $gc$ at time $t$.

\subsubsection{Post line contingencies\label{sec:lout}}
A line outage changes the topology of the system and therefore also the matrix $M$ defined in Sec. \ref{sec:normal}. Let $M^{\ell c}$ denote the matrix $M$ corresponding to the topology after a line outage $\ell c$. 
Using these notations, we model the line flow during a line contingency, $\ell c$, as follows:
\begin{flalign}
\boldsymbol f^{\ell c}_{\ell}(t) = M^{\ell c}_{(\ell,\cdot)} \left(\bar p(t) + \bar{\mu}(t) - \bar d(t) + \bar{\boldsymbol \omega}(t) - \boldsymbol \Omega(t) \bar \alpha(t)\right) \label{eq:randomlineflowcontingency}
\end{flalign}
where, $\boldsymbol f^{\ell c}_{\ell}(t)$ is the random line flow on $\ell$ during contingency $\ell c$ at time $t$. \linecomment{As discussed in the previous subsection}, unlike generator outages, \linecomment{the generator production levels are not modified in the case of line outages}.

\subsection{Optimization problem}\label{sec:optformulation}
With the notations and modeling considerations in Sec. \ref{sec:nomenclature} -- \ref{sec:lout}, 
we next present the formulation of the SCCUC. The objective function of the SCCUC minimizes the operating cost of the generators. The cost includes no-load cost, start-up cost, the running cost of all the generators and the cost of \linecomment{procuring reserve capacities}, i.e.
\begin{equation}
\begin{split}
\min \sum_{i\in \mathcal G} \sum_{t \in \mathcal T} \bigl\{a^0_i\cdot x_i(t) + \sum_{k\in \mathcal K} K_i^k \cdot g_i^k(t) + sc_i(t) +  \\
 \left[a^1_i\cdot (r^+_i(t)+ r^-_i(t)) + a^2_i\cdot r_i^{out}(t)\right] \bigr\} \label{eq:obj}
\end{split}
\end{equation}
The objective function is motivated by the models in \cite{Pandzic2015,Padhy2004,Wang2008}. \kaarthik{The optimization is subject to constraints in Eqs. (\ref{eq:randomlineflow})-(\ref{eq:randomlineflowcontingency}) and the following constraints}:

\noindent \textit{\underline{\smash{Binary variable logic:}}} The following set of constraints enforce the binary variable logic. 
% \subsubsection{Binary variable logic\label{sec:logic}}
\begin{subequations}
\begin{flalign}
&y_i(t) - z_i(t) = x_i(t) - x_i(t-1) \quad \forall t\in \mathcal T, i\in \mathcal G, \label{eq:logic1}\\
&y_i(t) + z_i(t) \leqslant 1 \quad \forall t \in \mathcal T, i\in \mathcal G. \label{eq:logic2}
\end{flalign}
\label{eq:logic}
\end{subequations}
Eq. \eqref{eq:logic1} determines if the generator is started up or shut down at hour $t$ based of its on-off status between hour $t$ and $t-1$. Eq. \eqref{eq:logic2} ensures start-up and shut-down of any generator do not occur at the same hour.  \\
\noindent \textit{\underline{\smash{Generation limits:}}} The deterministic and chance-constrained generation limits during normal operations and the reserve generation limits are enforced by the following set of constraints:
% \subsubsection{Generation limits\label{sec:genlimits}}
\begin{subequations}
\begin{flalign}
%&p_i^{\min} \cdot x_i(t) \leqslant p_i(t) \leqslant p_i^{\max} \cdot x_i(t) \quad \forall i \in \mathcal G, t\in \mathcal T, \label{eq:genlimits1} \\ 
&0 \leqslant r_i^-(t),r_i^+(t),r_i^{out}(t) \leqslant R_i \cdot x_i(t) \quad \forall i \in \mathcal G, t\in \mathcal T, \label{eq:genlimits2a} \\
&\delta_i^{gc}(t) \leqslant r_i^{out}(t) \quad \forall i \in \mathcal G, t\in \mathcal T, gc\in \mathcal C_g, \label{eq:genlimits2b} \\
%&\sum_{n\in \mathcal G} r_n^{out}(t) \geqslant p_i(t) \quad \forall i\in \mathcal G, t\in \mathcal T, \label{eq:genlimits2c} \\
&p_i(t) - r_i^-(t) \geqslant p_i^{\min} \cdot x_i(t) \quad \forall i \in \mathcal G, t\in \mathcal T, \label{eq:genlimits3} \\
&p_i(t) + r_i^+(t) + r_i^{out}(t) \leqslant p_i^{\max} \cdot x_i(t) \quad \forall i \in \mathcal G, t\in \mathcal T, \label{eq:genlimits4} \\
&\operatorname{Pr}(r_i^-(t) \geqslant \boldsymbol \Omega(t) \alpha_i(t)) \geqslant 1-\varepsilon_i \quad \forall i \in \mathcal G, t\in \mathcal T, \label{eq:genlimits5a} \\
&\operatorname{Pr}(r_i^+(t) \geqslant -\boldsymbol \Omega(t) \alpha_i(t)) \geqslant 1-\varepsilon_i \quad \forall i \in \mathcal G, t\in \mathcal T, \label{eq:genlimits5b} \\
%&\operatorname{Pr}(r_i^-(t) \geqslant \boldsymbol \Omega(t) \alpha_i^c(t)) \geqslant 1-\varepsilon_i \quad \forall i \in \mathcal G, t\in \mathcal T, c\in \mathcal C, \label{eq:genlimits6a} \\
%&\operatorname{Pr}(r_i^+(t) \geqslant -\boldsymbol \Omega(t) \alpha_i^c(t)) \geqslant 1-\varepsilon_i \quad \forall i \in \mathcal G, t\in \mathcal T, c\in \mathcal C, \label{eq:genlimits6b} \\
%&0 \leqslant \alpha_i(t) \leqslant x_i(t) \quad \forall i \in \mathcal G, t\in \mathcal T, \label{eq:participationa} \\
&0 \leqslant \alpha_i(t) \leqslant x_i(t) \quad \forall i \in \mathcal G, t\in \mathcal T. \label{eq:participation}
\end{flalign}
\label{eq:genlimits}
\end{subequations}
\linecomment{
The constraints in \eqref{eq:genlimits2a} -- \eqref{eq:genlimits4} enforce the generation limits and the reserve capacity limits for the generator $i$ at every hour $t$. 
Constraint \eqref{eq:genlimits2a} enforces an upper bound on the reserve capacities procured from a single generator, which may arise from, e.g., technical constraints on generator ramp rates. 
Constraint \eqref{eq:genlimits2b} ensures that the reserve capacity $r_i^{out}(t)$ procured from the generator $i$ during an hour $t$ must be greater than any reserve activation $\delta_i^{gc}$ expected from that generator. Note that \eqref{eq:powerbalancegout} in combination with \eqref{eq:genlimits2b} ensures that the total procured reserves are sufficient to cover any generator outage. 
The constraints \eqref{eq:genlimits3}, \eqref{eq:genlimits4} ensure that the minimum and maximum generation at generator $i$, defined as the combination of the scheduled generator output $p_i(t)$ and the minimum or maximum reserve activation (corresponding to full use of the reserve capacities), are within the generator limits $p_i^{min}, p_i^{max}$.  
The reserve constraints \eqref{eq:genlimits5a} -- \eqref{eq:genlimits5b} are \emph{chance constraints} on the reserve capacities $r_i^+(t), r_i^-(t)$. They guarantee, with a high probability $1-\varepsilon_i$, that we have sufficient generation reserves available to allow our generation control policy \eqref{eq:genresponse} to respond to wind fluctuations. Note that the violation probability $\varepsilon_i$ corresponds to the probability of requesting more reserves from a generator than we originally had procured.}
We do not model the system (emergency) response when we run out of reserve capacity, \linecomment{but allow the user to specify their willingness to accept the risk of reserve deficiency through} the user-specified parameter $\varepsilon_i$ for each generator $i$.
Constraint \eqref{eq:participation} imposes bounds on the participation factors of generators, \linecomment{based on their on-off status $x_i$}. \\
\noindent \textit{\underline{\smash{Production cost of the generators:}}} The \kaarthik{convex and} piece-wise linear production cost of each generator is defined by the following set of constraints:
\begin{subequations}
\begin{flalign}
&p_i(t) = \sum_{k\in \mathcal K} g_i^k(t) \quad \forall i \in \mathcal G, t\in \mathcal T, \label{eq:production1} \\
&0 \leqslant g_i^k(t) \leqslant p_{i,k}^{\max} \cdot x_i(t) \quad \forall i \in \mathcal G, t\in \mathcal T, k \in \mathcal K. \label{eq:production2}
\end{flalign}
\label{eq:prodcost}
\end{subequations}
Constraint \eqref{eq:production1} defines the power generated by each generator $i$ at each hour $t$ as the sum of power generated on each block of the production cost curve.  Constraint \eqref{eq:production2} enforces the limits on the power generated on each block. \\
\noindent \textit{\underline{\smash{Start-up cost of the generators:}}} The following set of constraints enforce the step-wise start-up cost for each generator:
\begin{subequations}
\begin{flalign}
&\sum_{s \in \mathcal S_i} w_{is}(t) = y_i(t) \quad \forall i \in \mathcal G, t\in \mathcal T, \label{eq:startup1}\\
&w_{is}(t) \leqslant \sum_{\underline T_{is}}^{\overline T_{is}} z_i(t-s) %+ \mathbf I(j < J \land j \leqslant t-1+\downinit<j+1) \notag\\
%&\qquad +\mathbf I(j=J\land j\leqslant t-1 +\downinit) 
\quad \forall i \in \mathcal G, t\in \mathcal T, s\in \mathcal S_i, \label{eq:startup2}\\
&sc_i(t) = \sum_{s\in \mathcal S_i} w_{is}(t) \cdot c_{is} \quad \forall i \in \mathcal G, t\in \mathcal T.\label{eq:startup3} 
\end{flalign}
\label{eq:startup}
\end{subequations}
The start-up cost for generator $i$ varies with the number of consecutive time periods it has been switched off before it is started up. Constraint \eqref{eq:startup1} ensures exactly one start-up cost from the set of start-up cost blocks is chosen for generator $i$. Constraint \eqref{eq:startup2} identifies the appropriate start-up block by implicitly counting the number of consecutive time periods the generator has been in the off state. Finally, constraint \eqref{eq:startup3} selects the actual start-up cost for the objective function. \\
\noindent \textit{\underline{\smash{Minimum up-time, minimum down-time, and ramping:}}}
\begin{subequations}
\begin{flalign}
&x_i(t) =  p_i^{\operatorname{on-off}} \quad \forall i\in \mathcal G, t\leqslant \overline L_i + \underline L_i, \label{eq:updown1} \\
&\sum_{n=\overline{t}}^t y_i(n) \leqslant x_i(t) \quad \forall i\in \mathcal G, t\geqslant \overline L_i, \overline{t} = t-UT_i+1, \label{eq:updown2} \\
&\sum_{n=\underline{t}}^t z_i(n) \leqslant 1- x_i(t) \quad \forall i\in \mathcal G, t\geqslant \underline L_i, \underline{t} = t-DT_i+1, \label{eq:updown3} \\
&RD_i \geqslant p_i(t-1) - p_i(t) \quad \forall i \in \mathcal G, t\in \mathcal T, \label{eq:rampdown}\\
&RU_i \geqslant p_i(t) - p_i(t-1) \quad \forall i \in \mathcal G, t\in \mathcal T. \label{eq:rampup}
\end{flalign}
\label{eq:updownramp}
\end{subequations}
The constraint in Eq. \eqref{eq:updown1} sets the on-off status of generator $i$ based on initial conditions. Notice that both $\underline L_i$ and $\overline L_i$ will not take positive values simultaneously. The constraints in Eqs. \eqref{eq:updown2} and \eqref{eq:updown3} enforce minimum up-time and minimum down-time constraints for generator $i$ for the remaining time intervals of the planning horizon. The constraints \eqref{eq:rampdown} and \eqref{eq:rampup} enforce ramping limits on consecutive periods on every generator $i$. \\
\noindent \textit{\underline{\smash{Power flow:}}}
\begin{subequations}
\begin{flalign}
%&\sum_{i\in \mathcal G} \alpha_i^c(t) = 1, \alpha_r^c(t) = 0  \quad \forall t \in \mathcal T, c \in \mathcal C,\label{eq:powerflow1b}\\ 
%&\alpha_r^c(t) = 0 \quad \forall t \in \mathcal T, c \in \mathcal C \label{eq:powerflow2b} \\
&\sum_{i\in \mathcal G} \alpha_i(t) = 1, \alpha_r(t) = 0  \quad \forall t \in \mathcal T, \label{eq:powerflow1a}\\ 
%&\alpha_r(t) = 0 \quad \forall t \in \mathcal T, \label{eq:powerflow2a} \\
&\sum_{b\in \mathcal B} (p_b(t) + \mu_b(t) - d_b(t)) = 0 \quad \forall t\in \mathcal T, \label{eq:powerflow3a}\\
&\sum_{i\in \mathcal G} \delta^{gc}_i(t) = 0 \text{ and } \delta^{gc}_{gc}(t) = - p_{gc}(t) \quad \forall t \in \mathcal T, gc \in \mathcal C_g, \label{eq:powerflow3b}\\
&\operatorname{Pr}(\boldsymbol f_{\ell}(t) \leqslant f^{\max}) > 1- \varepsilon_{\ell} \quad \forall \ell \in \mathcal L, t\in \mathcal T,  \label{eq:powerflow4a}\\
&\operatorname{Pr}(\boldsymbol f_{\ell}(t) \geqslant -f^{\max}) > 1- \varepsilon_{\ell} \quad \forall \ell \in \mathcal L, t\in \mathcal T,  \label{eq:powerflow4b}\\
&\operatorname{Pr}(\boldsymbol f^{\ell c}_{\ell}(t) \leqslant f^{\max}) > 1- \varepsilon_{\ell}^{\ell c} \quad \forall \ell \in \mathcal L, t\in \mathcal T, {\ell c}\in \mathcal C_{\ell}, \label{eq:powerflow5a}\\
&\operatorname{Pr}(\boldsymbol f^{\ell c}_{\ell}(t) \geqslant -f^{\max}) > 1- \varepsilon_{\ell}^{\ell c} \quad \forall \ell \in \mathcal L, t\in \mathcal T, \ell c \in \mathcal C_{\ell}, \label{eq:powerflow5b} \\
& -f^{\max} \leqslant f_{\ell}^{gc}(t) \leqslant f^{\max} \quad \forall \ell \in \mathcal L, t\in \mathcal T,  gc \in \mathcal C_{g}. \label{eq:powerflow6}
\end{flalign}
\label{eq:powerflow}
\end{subequations}
The power flow equations are adapted from \cite{Bienstock2014, Lubin2015, Roald2016} and are generalized for multiple time periods. Constraint \eqref{eq:powerflow1a} guarantees balance of generation and load via the participation factors (Sec. \ref{sec:gencontrol}). Constraints \eqref{eq:powerflow3a} and \eqref{eq:powerflow3b} impose the demand-generation balance during normal operations and contingencies for all time periods. Finally, constraints \eqref{eq:powerflow4a} -- \eqref{eq:powerflow5b} are the chance constraints for the line flows during normal operations and during line contingencies. $\varepsilon_{\ell}$ and $\varepsilon_{\ell}^{\ell c}$ are user-defined parameters to control to probability of line violations during normal operations and post line contingencies, respectively. Constraint \eqref{eq:powerflow6} imposes hard line limits on the lines for post generator contingencies; $\boldsymbol f_{\ell}$ and $\boldsymbol f_{\ell}^{\ell c}$ are computed using equations \eqref{eq:randomlineflow} and \eqref{eq:randomlineflowcontingency} respectively.

\subsection{Reformulation of Chance Constraints} \label{subsec:reformulation}
The chance constraints in \eqref{eq:genlimits5a}, \eqref{eq:genlimits5b}, \eqref{eq:powerflow4a} - \eqref{eq:powerflow5b} are often non-convex and difficult to handle. But, under the assumption of normality, these equations can be reformulated to a convex form that is computationally tractable (see \cite{Bienstock2014, Lubin2015, Roald2013}). In particular, if $\bm \xi$ is a multi-variate normal random variable with mean $\mu$ and covariance $\Sigma$ and if $v$ is the decision variable vector, then a chance constraint of the form 
\begin{flalign}
& \operatorname{Pr}(\bm \xi^T v \leqslant b) \geqslant 1-\varepsilon \label{eq:cc}
\end{flalign}
is equivalent to 
\begin{flalign}
& \mu^T v + \phi^{-1} (1-\varepsilon) \sqrt{v^T\Sigma v} \leqslant b. \label{eq:reformulation}
\end{flalign}
In \eqref{eq:reformulation}, $\phi^{-1}$ denotes the inverse cumulative distribution function of the standard normal distribution. The constraint \eqref{eq:reformulation} is a convex, SOC constraint. We also remark that the chance constraints in \eqref{eq:genlimits5a} and \eqref{eq:genlimits5b} contain a special structure that enables reformulation to a linear constraint. 
\begin{proposition} \label{prop:lin}
The chance constraints in Eqs. \eqref{eq:genlimits5a} and \eqref{eq:genlimits5b} permit a linear reformulation. 
\end{proposition}
\begin{proof}
Consider the chance constraint in \eqref{eq:genlimits5a} given by 
\begin{flalign*}
&\operatorname{Pr}(r_i^-(t) \geqslant \boldsymbol \Omega(t) \alpha_i(t)) \geqslant 1-\varepsilon_i.
% \quad \Rightarrow \quad \operatorname{Pr}(r_i^-(t) - \boldsymbol \Omega(t) \alpha_i(t)\geqslant 0) \geqslant 1-\varepsilon_i.
\end{flalign*}
In this constraint, $\bm \Omega(t)$ is the total deviation from the forecast which is a \linecomment{scalar} Gaussian random variable with mean $\mu_\Omega = \sum_{b\in \mathcal{W}} \mu_b(t)$ and variance $\sigma_\Omega = e^T \Sigma(t) e$. Here, $e$ denotes a vector of ones and 
\linecomment{$\Sigma(t)$ is the covariance matrix of the random vector $\bar \omega(t)$, which may include non-zero off-diagonal elements to represent correlated random variables}.
%\kaarthik{$\Sigma(t)$ is a diagonal covariance matrix with entries $\sigma_b(t)^2$, $b \in \mathcal W$}. 
Using the above notations and Eq. \eqref{eq:reformulation}, the chance constraint is equivalent to the following linear constraint:
\begin{flalign}
& r_i^-(t) - \alpha_i(t)\sum_{b \in \mathcal W} \mu_b(t) + \phi^{-1}(1-\varepsilon_i) \alpha_i(t)\sqrt{e^T \Sigma(t) e} \leqslant 0. \label{eq:linearcc}
\end{flalign}
A similar reduction holds for the chance constraint in \eqref{eq:genlimits5b}. 
\end{proof}

\subsection{Additional valid inequalities} \label{subsec:vi}
In this section, we present some additional valid inequalities that strengthen the continuous relaxation of the formulation presented in \ref{sec:optformulation}. \kaarthik{The following result presents strengthened versions of the ramping constraints in Eq. \eqref{eq:rampdown} and \eqref{eq:rampup}}.
\begin{proposition} \label{prop:ramp_1}
The following inequalities are valid for the SCCUC and strengthen the constraints in Eqs. \eqref{eq:rampdown} and \eqref{eq:rampup}
\begin{subequations}
\begin{flalign}
&RD_i \left( x_i(t) + z_i(t) \right)  \geqslant p_i(t-1) - p_i(t) \quad \forall i \in \mathcal G, t\in \mathcal T, \label{eq:rampdown_1}\\
&RU_i \left(x_i(t-1) + y_i(t)\right) \geqslant p_i(t) - p_i(t-1) \quad \forall i \in \mathcal G, t\in \mathcal T. \label{eq:rampup_1}
\end{flalign}
\label{eq:ramp_1}
\end{subequations}
\end{proposition}
\begin{proof}
The claim that the equations in \eqref{eq:ramp_1} strengthen the respective constraints in Eqs. \eqref{eq:rampdown} and \eqref{eq:rampup} is trivial to observe. Validity of Eq. \eqref{eq:ramp_1} is also easy to observe under the trivial scenarios when, for a given $i\in \mathcal G$ and $t \in \mathcal T$, $x_i(t) = x_i(t-1) = 1$; in these scenarios the equations in Eq. \eqref{eq:ramp_1} reduce to the ramping constraints in Eq. \eqref{eq:rampdown} and \eqref{eq:rampup}. The validity of Eqs. \eqref{eq:rampdown_1} and \eqref{eq:rampup_1} for the other nontrivial scenarios can be seen by examining the constraints under the following two cases for a given $i \in \mathcal G$ and $t\in \mathcal T$: (i) $x_i(t-1) = 0$, $x_i(t) = y_i(t) = 1$, $z_i(t) = 0$, and (ii) $x_i(t-1) = 1$, $x_i(t) = z_i(t) = 1$, $y_i(t) = 0$. For case (i), the equations in \eqref{eq:ramp_1} reduce to 
\begin{flalign} 
p_i(t) \geqslant -RD_i \text{ and }  p_i(t) \leqslant RU_i(t), \label{eq:ramp_case_1}
\end{flalign}
due to the constraints \eqref{eq:genlimits2a}, \eqref{eq:genlimits3}, and \eqref{eq:genlimits4}; Eq. \eqref{eq:ramp_case_1} are trivially satisfied by any solution to the SCCUC. Similarly, case (ii) can be argued, completing the proof. 
\end{proof}
These additional valid inequalities are added as additional constraints to optimization problem presented in Sec. \ref{sec:optformulation}.
\section{Solution Methodology} \label{sec:algo}
In this section, we present our modified Benders decomposition algorithm to solve the SCCUC problem. 
For the purposes of our algorithm, we classify the constraints into three categories:

\begin{enumerate}[label=(\Alph*)]

\item  Normal operation, unit commitment constraints: Constraints \eqref{eq:logic1} -- \eqref{eq:genlimits2a}, \eqref{eq:genlimits3}, \eqref{eq:genlimits5a} -- \eqref{eq:powerflow3a}, \eqref{eq:powerflow4a},\eqref{eq:powerflow4b}, and \eqref{eq:ramp_1},

\item Line contingency constraints: Constraints \eqref{eq:powerflow5a} and \eqref{eq:powerflow5b}

\item Generator contingency constraints: Constraints \eqref{eq:genlimits2b}, \eqref{eq:genlimits4}, \eqref{eq:powerflow3b}, and \eqref{eq:powerflow6}. 

\end{enumerate}

\noindent
The resulting SCCUC problem is a large, MISOCP. Even large-scale continuous SOC problems are often difficult for commercial solvers (despite their nice theoretical properties). Hence, we use a sequential outer-approximation technique to handle the SOC constraints. This approach was effective on similar problems in transmission systems (see \cite{Bienstock2014} and \cite{optimal_soc} for SOCP and MISOCP, respectively).

\subsection{Modified Benders decomposition} \label{subsec:Benders}
 While a Generalized Benders decomposition algorithm based on SOCP duality exists for the SCCUC (see \cite{Sundar2016}), the approach has convergence and numerical issues for even medium sized test instances. The  convergence issues are due to the line contingency chance constraints in category (B); the number of line contingency chance constraints is $(|\mathcal L|-1) \times |\mathcal L| \times |\mathcal T|$. \cite{Sundar2016} formulated the constraints in category (B) as Benders feasibility sub-problems and generated feasibility cuts using SOC duality theory. The main challenge with this approach is that current state-of-the-art SOC programming solvers are not mature enough to give reliable extreme rays for infeasible SOC programs. We remark that this is just a practical implementation issue and not a theoretical one. The  alternative is to add all the constraints in category (B) to the Benders master problem and perform sequential outer-approximation on these constraints to solve the master problem. This results in a large number of SOC constraints in a Benders master problem, thereby (excessively) increasing the computation time to solve the master problem. Furthermore, it has been observed that for power-flow based problems in transmission systems and in the presence of wind, the number of SOC constraints in category (B) that actually are tight for the optimal solution are very minimal \cite{Bienstock2014, Roald2016}. Based on these observations, we were motivated to develop a modification to the Benders decomposition algorithm for problems where the \textit{master problem contains a large number of SOC constraints of which very few are binding}.

% {\color{red} We should refer to Algorithm 1 with line numbers within the next few paragraphs}

 Now, we present a high-level overview of the approach and, in the following sections, we formalize the algorithm. This technique first relaxes the line contingency constraints in category (B) from the SCCUC to obtain a relaxed SCCUC (R-SCCUC) formulation. This R-SCCUC is then decomposed into a MISOC master problem and linear programming sub-problems. The master problem contains all the constraints in category (A) reformulated as SOC constraints (Step \ref{step:outer} of algorithm \ref{algo:pseudocode}). Each sub-problem spans a single time period and consists of all the generator contingency constraints from category (C) for that time period, with the unit commitment variables and generator dispatch variables during normal operating conditions fixed to values obtained from solving the master problem. This relaxed problem (R-SCCUC) is solved using the traditional Benders decomposition algorithm (Step \ref{step:outer} -- \ref{step:endbenders} of algorithm \ref{algo:pseudocode}). 
 
 We next present an overview of the Benders decomposition algorithm in order to keep the presentation self-contained. The optimal solution to the master problem determines the running, start-up and shut-down schedule of the generators with their dispatch during normal operating conditions. For the schedule fixed from the master problem, the Benders sub-problems determine the \linecomment{total generation reserve capacity necessary to cover generator outages}. The solutions to the sub-problem provide information on the quality of the dispatch decisions made by the master problem. This information is passed to the master problem again as Benders optimality or feasibility cuts. 
%for the infeasible and optimal sub-problems, respectively. 
The cuts enable the master problem to modify and compute better dispatch and generation levels. This iterative procedure continues until the Benders decomposition termination condition is satisfied. As for line contingency constraints in category (B), the optimal solution obtained for the R-SCCUC problem via the Benders decomposition is verified to see if it violates any of the these constraints. If so, these infeasible SOC constraints are added to the master problem of the R-SCCUC and the resulting problem is resolved again using the Benders decomposition algorithm. The pseudocode of the modified Benders decomposition algorithm is given in  Algorithm \ref{algo:pseudocode}. We note that the MISOC master problem in the Benders decomposition algorithm is solved via sequential outer-approximation of the SOC constraints. 

\begin{proposition}
The modified Benders decomposition algorithm, whose pseudocode is given in algorithm \ref{algo:pseudocode}, converges to an optimal solution of the SCCUC in finite number of iterations.
\end{proposition}
\begin{proof}
The proof of finite convergence is argued in two steps. In the first step, we claim that the inner loop of the algorithm \ref{algo:pseudocode} converges to an optimal solution of the problem in the inner loop (Step \ref{step:outer} -- \ref{step:endbenders} of algorithm \ref{algo:pseudocode}) in finite number of iterations. This is trivial to observe because the inner loop solves a MISOC using traditional Benders decomposition with linear programming subproblems. Hence, the convergence and termination properties of the traditional Benders decomposition algorithm extends to the MISOC in the inner loop. As for the outer loop of the modified Benders algorithm which solves the R-SCCUC, it terminates when all the category (B) constraints are feasible for the current solution to the R-SCCUC. Since there are only finite number of category (B) constraints, even in the worst case, the outer loop would terminate in finite number of iterations to an optimal solution of the SCCUC. 
\end{proof}

% {\color{red} Its obvious, but we should include a proof of convergence for the modified Benders... }

In summary, the modified Benders decomposition algorithm relaxes a large number of SOC constraints (category (B) constraints) from the SCCUC and solves the resulting R-SCCUC using a combination of sequential outer-approximation and traditional Benders decomposition. The relaxed infeasible SOC constraints is then added back to the R-SCCUC in an iterative fashion with the traditional Benders decomposition in combination with the sequential outer-approximation being used to solve the R-SCCUC during each iteration. 
%We now present a pseudo code of the algorithm followed by the the sub-problem and the master problem for the Benders decomposition. 

{
\begin{algorithm}[h!]
\caption{Modified Benders Decomposition - Pseudo Code}\label{algo:pseudocode}
\small
\onehalfspacing
\begin{algorithmic}[1]
\Loop
\Comment {Outer loop: R-SCCUC} %Relax category (B) constraints to obtain R-SCCUC}
\State \label{step:outer} Solve master problem via sequential outer-approximation
\Loop 
\State Fix master problem variables \& solve the sub-problems
\State Generate feasibility and optimality cuts
\State Resolve master problem with the additional cuts
\State Check Benders decomposition termination condition
\EndLoop \label{step:endbenders}
\If{All the category (B) constraints are feasible}
\State Exit with optimal solution for SCCUC
\Else 
\State Augment violated category (B) constraints to R-SCCUC  \label{step:cc}
\State Re-solve (go to step 2).
\EndIf
\EndLoop
\end{algorithmic}
\end{algorithm}
}

\subsection{Benders sub-problems} \label{subsec:sub}
The Benders sub-problems are formulated with the unit commitment decisions and the generation level solutions of the master problem and they compute new generation levels for each generator outage. One sub-problem is formulated for each time period and the objective of each sub-problem minimizes the total \linecomment{generator reserve capacity $r_i^{out}(t)$ allocated to cover all generator outages} during that time period. For a fixed time period $t$, the sub-problem is as follows:
\begin{flalign}
&SP(t, \bar x(t), \bar r^+(t), \bar p(t)): \min \sum_{i \in \mathcal G} a_i^2 r^{out}_i(t) \, \mathrm{subject \ to} \label{eq:sp-obj} \\
&0 \leqslant r_i^{out}(t) \leqslant R \cdot x_i(t) \quad \forall i \in \mathcal G, \notag  \\
&\delta_i^{gc}(t) - r_i^{out}(t) \leqslant 0 \quad \forall i \in \mathcal G, gc\in \mathcal C_g, \notag \\
&\sum_{n\in \mathcal G} r_n^{out}(t) \geqslant p_i(t) \quad \forall i\in \mathcal G, \notag \\
&r_i^{out}(t) \leqslant p_i^{\max}\cdot x_i(t) - p_i(t) - r_i^+(t)  \quad \forall i \in \mathcal G, \notag \\
&p^{gc}_i(t) - \delta^{gc}_i(t) = p_i(t) \quad \forall i \in \mathcal G, gc \in \mathcal C_g, \notag \\
& \sum_{i\in \mathcal G} \delta^{gc}_i(t) = 0, \text{ and } \delta^{gc}_{gc}(t) = - p_{gc}(t)  \quad \forall gc \in \mathcal C_g, \notag \\
& -f^{\max} \leqslant f_{\ell}^{gc}(t) \leqslant f^{\max} \quad \forall \ell \in \mathcal L,  gc \in \mathcal C_{g}. \notag 
%& \text{limits on $\bar r^{out}(t)$ in \eqref{eq:genlimits2a} -- \eqref{eq:genlimits2c}, \eqref{eq:genlimits4}}, \notag & \\
%& \text{power balance for each generator outage \eqref{eq:genresponsecontingency} and \eqref{eq:powerbalancegout}, and} \notag & \\
%& \text{line limits for each generator outage \eqref{eq:powerflow6}}. \notag &
\end{flalign}

For ease of exposition and clarity, the sub-problem constraints shown in \eqref{eq:sp-obj} have the master problem variables on the \kaarthik{left-hand side}. Each sub-problem is a linear program and the values of the variables $\bar x(t)$, $\bar r^+(t)$ and $\bar p(t)$ are obtained from the master problem solution. A Benders feasibility or optimality cut is generated based on whether the sub-problem is infeasible or optimal, respectively. If we model the constraints of the sub-problem with ${A}(t){X}(t) \leqslant {b}(t)$, then the Benders optimality cut is given by $\eta(t) \geqslant {\bm \gamma}(t)^T b(t)$, where ${\bm \gamma}(t)$ are the dual values of the constraints. Similarly, the the feasibility cut is given by ${\bm \beta}(t)^T b(t) \leqslant 0$, where ${\bm \beta}(t)$ is the infeasible ray found using Farkas' lemma. 
These cuts are added to the master problem and the master problem is re-solved (see step 5 and 6 of algorithm \ref{algo:pseudocode}). The master problem objective value provides a lower bound and an upper bound is computed using the solution to the sub-problems. The inner loop of Algorithm \ref{algo:pseudocode} terminates when the gap between the upper and lower bound is within a tolerance value. We remark that during each iteration of the inner loop, $|\mathcal T|$ sub-problems are solved to generate the optimality and feasibility cuts. 

\subsection{Benders master problem} \label{subsec:master}
The Benders master problem minimizes the operation costs and generates a schedule of the generators with their power levels during normal operating conditions, i.e. %The master problem formulation is given below:
\begin{equation}
\begin{split}
\min & \sum_{i\in \mathcal G} \sum_{t \in \mathcal T} \bigl\{a^0_i\cdot x_i(t) + \sum_{k\in \mathcal K} K_i^k \cdot g_i^k(t) + sc_i(t) +  \\
& \qquad \left[a^1_i\cdot (r^+_i(t)+ r^-_i(t)) \right] \bigr\} + \sum_{t\in \mathcal T} \eta(t) \label{eq:mp-obj}
\end{split}
\end{equation}
subject to:
\begin{flalign}
% &\min \sum_{i\in \mathcal G} \sum_{t \in \mathcal T} \bigl\{a^0_i\cdot x_i(t) + \sum_{k\in \mathcal K} K_i^k \cdot g_i^k(t) + sc_i(t) + 
% \left[a^1_i\cdot (r^+_i(t)+ r^-_i(t)) \right] \bigr\} + \sum_{t\in \mathcal T} \eta(t) \label{eq:mp-obj} \\
% &\mathrm{subject \ to} \nonumber \\  
&\text{\eqref{eq:logic1} -- \eqref{eq:genlimits2a}, \eqref{eq:genlimits3}, \eqref{eq:genlimits5a} -- \eqref{eq:powerflow3a}, \eqref{eq:powerflow4a}, \eqref{eq:powerflow4b}, and \eqref{eq:ramp_1}}, \notag  \\
% &f^{\ell c}_{\ell}(t) = M^{\ell c}_{(\ell,\cdot)} \left(\bar p(t) + \bar{\mu}(t) - \bar d(t) \right) \quad \forall \ell \in \mathcal L, t\in \mathcal T,  \ell c \in \mathcal C_{\ell}, \label{eq:detlclineflows}  \\
%  &-f^{\max} \leqslant f_{\ell}^{\ell c}(t) \leqslant f^{\max} \quad \forall \ell \in \mathcal L, t\in \mathcal T,  \ell c \in \mathcal C_{\ell}. \label{eq:detlinecontingencies} %\\
& \eta(t) \geqslant \bm \gamma(t)^T  b(t) \quad \forall t \in \mathcal T\label{eq:optimality}, \text{ and} \\
&  {\bm \beta}(t)^T  b(t) \leqslant 0 \quad \forall t \in \mathcal T\label{eq:feasibility}.
\end{flalign}
The variables $\eta(t)$ are surrogate variables introduced in the master problem for each sub-problem. They globally underestimate the value of the generation \linecomment{outage} reserves, $\sum_i r^{out}_i(t)$, for each $t$. The equations \eqref{eq:optimality} and \eqref{eq:feasibility} are the optimality and feasibility cuts generated using the dual value, $\boldsymbol{\gamma}(t)$, and infeasibility ray, $\boldsymbol{\beta}(t)$, for the optimal and infeasible Benders sub-problems.

\section{Computational results} \label{sec:casestudy}
In this section, we demonstrate the computational effectiveness of the modified Benders decomposition algorithm to solve the SCCUC \linecomment{on two test systems, the three area IEEE RTS-96 system \cite{Wong1999} and the NESTA IEEE 300 system \cite{Coffrin2014}}. We then present the benefits of solving the SCCUC relative to the deterministic version of the unit commitment problem using the three area IEEE RTS-96 system. The comparison between the SCCUC and its deterministic counterpart is based on a variety of factors including nominal operating cost, number of generator and line violations, and amount of \linecomment{reserve capacities allotted}. All computational experiments were run on a Intel Haswell 2.6 GHz, 62 GB, 20-core super computing machine at Los Alamos National Laboratory.

\subsection{Test system and wind data}
\begin{figure}[htbp]
\centering
\begin{tikzpicture}[scale=0.6]
\begin{axis}[
  xlabel=Hour,
  ylabel=Power (MW),
  legend style={at={(-0.1,1.1)},anchor=west,draw=none},
  legend columns=-1]
\addplot +[mark=none, line width=2pt] table [y=Load, x=Time]{wind-loaddata};
\addlegendentry{Load}
\addplot +[mark=none, line width=1pt] table [y=Wind3, x=Time]{wind-loaddata};
\addlegendentry{Favourable wind}
\addplot +[dashed, mark=none, line width=1pt] table [y=Wind2, x=Time]{wind-loaddata};
\addlegendentry{Unfavourable wind}
\end{axis}
\end{tikzpicture}
\caption{Load profile and mean wind power injection data}
\label{fig:data}
\end{figure}
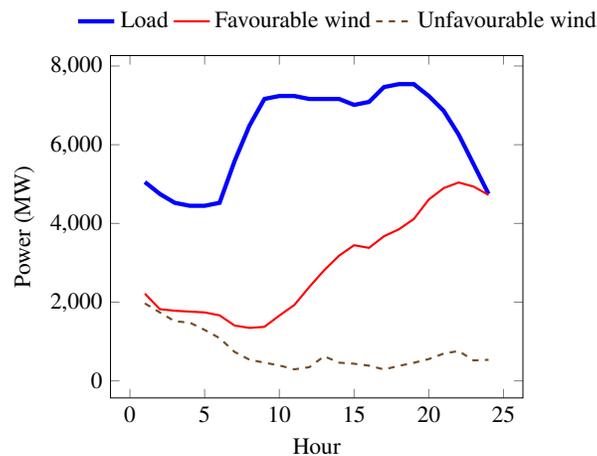

\linecomment{
\subsubsection{IEEE RTS-96 System}
}
The IEEE RTS-96 \cite{Wong1999}, modified to accommodate $19$ wind farms, was used for empirical testing. The system consists of 73 buses, 120 transmission lines and 96 conventional generators. The total installed capacity of the generators is 10,215 MW. Among the 96 generators, 6 are nuclear and 3 are hydro generators. The NREL Western Wind dataset \cite{Potter2008} provides the wind data. Wind farms locations are mapped to the IEEE RTS-96 respecting the lengths of the lines (see \cite{Pandzic2015}). The test system contains a total of 19 wind farms with a total generation capacity of 6900 MW. The locations of the wind farms, the individual generation capacity of each wind farm, the overall system diagram; the stepwise generation cost, start-up cost, ramping restrictions, up and down-time restrictions for each generator; and the load profile data for a 24-hour period used for the case study are in \cite{Pandzic2015} and can be found at \url{http://www.ee.washington.edu/research/real/library.html}. We use hourly wind forecasts for two sets of 5-day periods for each wind farm provided in \cite{Pandzic2015}. Using this forecast data for each 5-day period, the mean and standard deviation of the wind power injection for each time period was estimated assuming that the wind power injections for each hour are independent normally distributed random variables. We classify the wind data sets as favourable and unfavourable; favourable wind data indicates wind power injections increasing with the total load during each time period, \linecomment{while unfavourable indicates that the wind power injections move in the opposite direction of the total load}. Figure \ref{fig:data} shows the total load profile and the sum of the mean wind power injection at the wind farms for both the wind data sets. The cost coefficients $a_i^0,~a_i^1,$ and $a_i^2$ for the generation and tertiary reserves, respectively, are adopted directly from the IEEE RTS-96 generation cost coefficient data and wind power is assumed to have zero marginal cost.  

\linecomment{
\subsubsection{NESTA IEEE 300 System}
Generally, there is a lack of established large test systems for the SCCUC, since, even small problems are hard to solve. Thus, in order to test the scalability of the algorithm for large test cases, we created a modified version of the NESTA IEEE 300 network found in \cite{Coffrin2014}. 
%Kaarthik, can we add a few sentences describing the data for the NESTA IEEE 300 bus system here? I think that would be the most efficient way to address the Reviewer 2, Question 6?
}

\subsection{Solving the MISOCP}
The MISOCP presented in Sec. \ref{sec:formulation} for the SCCUC can be directly solved by off-the-shelf commercial solvers like CPLEX or Mosek. Initial numerical simulations indicate that the MISOCP is challenging to solve directly on the three area IEEE RTS-96 test system. For instance, CPLEX 12.7 \cite{cplex} was not able to produce any feasible solution after running for $2$ hours. After presolving the problem in CPLEX, the problem contained 7,123,102 rows, 6,588,222 columns and 184,559,537 nonzeros, 6,273 binaries and 334,920 quadratic constraints. \kaarthik{We remark that we did not attempt to solve the MISOCP for the NESTA IEEE 300 test system using CPLEX as it already had difficulties in producing even a feasible solution to the IEEE RTS-96 system}. This also motivates the need for a specialized algorithm like the modified Benders decomposition algorithm to solve the SCCUC.

\subsection{Performance of modified Benders decomposition algorithm}
\begin{figure}[htbp]
\centering
  \includegraphics[scale=0.45]{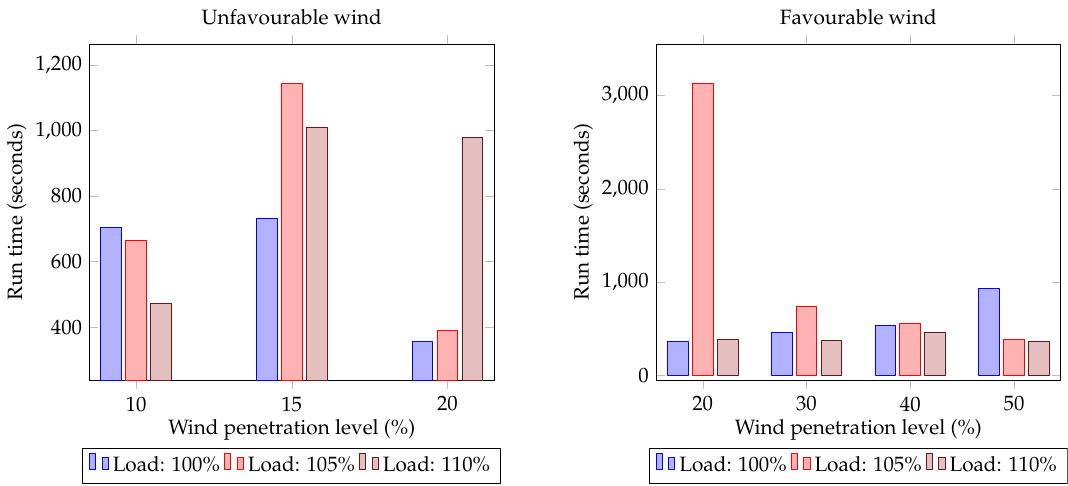}
  \caption{Computation times for the modified Benders algorithm}
  \label{fig:times}
\end{figure}

The SCCUC formulation has three user-defined parameters: $\varepsilon_i,~\varepsilon_{\ell},$ and $\varepsilon_{\ell}^{\ell c}$. For the rest of the computational experiments, we set them to $1\%,~10\%$ and $20\%$, respectively. 
Based on recent literature \cite{CIGRE2009}, many power system operators \kaarthik{want the probability of generator production limit and transmission line thermal limit violations to be less than $1\%$ and $10\%$, respectively, during normal operations. In the case of emergency (an $N-1$ failure is an example of an emergency), they permit the transmission lines thermal limit violation probability to be at most $20\%$}. 
We then sequentially vary the loading levels and solve the resulting SCCUC instances using the modified Benders decomposition algorithm in Sec. \ref{sec:algo}. The nominal load obtained from the data set is assumed to be 100\% and the loads are increased by 5\% and 10\% to stress the system further. For each loading level, both the favourable (F) and unfavourable (UF) wind data sets are used with different penetration rates. For all the runs, the optimality tolerance was set to 1\% and the number of threads was set to 4. The algorithms were implemented using the Julia programming language with the JuMP and JuMPChance modeling framework \cite{lubin2015computing} with CPLEX 12.7 as the commercial solver. The computation times are shown in Figure \ref{fig:times}. For all the runs, the modified Benders decomposition algorithm was terminated when the optimality gap was less than 1\%. From the figure, it is clear that the algorithm is very effective and is able to provide a feasible solution within 1\% of the best lower bound in less than an hour for all the runs. The actual computation times for all the runs is also shown in Tables \ref{tab:fav_convergence} -- \ref{tab:unfav_convergence}.

To corroborate the effectiveness of the algorithm from a scalability perspective, we ran the algorithm on an IEEE 300 NESTA test system in \cite{Coffrin2014} that consists of 300 buses, 411 lines, and 57 generators. The wind farm locations, means and variances for each wind farm were generated as follows: all the loads between 80 to 300 MW are assumed to be a wind farm that draws power with a standard deviation equal to 10\% of the load at that bus. This results in 51 wind farms. The number of generator and line contingencies in the system is 52 and 411, respectively. 5 generator contingencies are not feasible in this model and were removed. The other unit commitment data that is not available from the IEEE 300 test system is adapted from IEEE RTS-96 system.  \kaarthik{The overall run time of the algorithm to converge to a dispatch that is N-1 secure and is within 20\%, 10\%, 5\%, and 1\% of the optimal dispatch is 32 minutes, 53 minutes, 2 hours and 5 minutes, and 3 hours and 25 minutes, respectively.} 

Next, we present the convergence properties of the algorithm.  \kaarthik{Tables \ref{tab:fav_convergence} 
-- \ref{tab:unfav_convergence} present the number of iterations, given by \#I, of the outer loop of the modified Benders decomposition algorithm (see algorithm \ref{algo:pseudocode}) and, \#CC-V, the  number of lines for which the post-contingency line flow constraints were added back to the inner problem (in line \ref{step:cc} of Algorithm \ref{algo:pseudocode}) during the course of executing the algorithm.}
This table provides strong evidence of the effectiveness of our approach, as very few lines violate their limits. 
We remark that the line flow chance constraints for a single line can become infeasible in different time periods; in this case the violation on that line, as presented in the table \ref{tab:fav_convergence} 
-- \ref{tab:unfav_convergence}, is still counted only once. The tables indicate that the modified Benders decomposition is a very effective way of handling chance constraints on the line flow limits during line contingencies, via the outer loop of the algorithm, and converges in a few iterations. It also helps identify the lines that frequently overload in the presence of wind. For instance, the line contingency chance constraints on the lines $(107,108)$ and $(108,109)$ in the IEEE RTS-96 were violated after the first iteration of the algorithm \ref{algo:pseudocode} during every run. 

\begin{table}[htbp]
    \centering
        \caption{Number of lines \kaarthik{(out of 117) for which the post-contingency line flow chance constraints had to be added to the inner problem (\#CC-V),}
        number of iterations (\#I) of the modified Benders decomposition algorithm, and the computation time (seconds) for the modified Benders decomposition algorithm on the instances favourable (F) wind penetration rates.}
    \label{tab:fav_convergence}
    \footnotesize
    \setlength\tabcolsep{3.5pt}
    \begin{tabularx}{\columnwidth}{c|ccc|ccc|ccc}
    % \begin{tabular}{c|ccc|ccc|ccc}
    \toprule
        $W\%$ & \multicolumn{3}{c|}{Load$=100\%$} & \multicolumn{3}{c|}{Load$=105\%$} & \multicolumn{3}{c}{Load$=110\%$} \\
    \cmidrule(lr){2-10}
    F & \#CC-V & \#I & time & \#CC-V & \#I & time & \#CC-V & \#I & time \\
    \midrule
    10 & 3 & 2 & 703.47 & 3 & 2 & 665.57 & 3 & 2 & 472.49 \\
    15 & 5 & 2 & 732.82 & 3 & 2 & 1144.37 &  3 & 2 & 1007.79 \\
    20 & 5 & 2 & 355.45 & 3 & 2 & 388.62 & 3 & 2 & 977.88 \\
    \bottomrule
    \end{tabularx}
    % \end{tabular}
\end{table}

\begin{table}[htbp]
    \centering
        \caption{Number of lines \kaarthik{(out of 117) for which the post-contingency line flow chance constraints had to be added to the inner problem (\#CC-V),}
        %on which contingency line flow chance constraint violations (\#CC-V) 
        number of iterations (\#I) of the modified Benders decomposition algorithm, and the computation time (seconds) for the modified Benders decomposition algorithm on the instances with unfavourable (UF) wind penetration rates.}
    \label{tab:unfav_convergence}
    \footnotesize
    \setlength\tabcolsep{3.5pt}
    \begin{tabularx}{\columnwidth}{c|ccc|ccc|ccc}
    % \begin{tabular}{c|ccc|ccc|ccc}
    \toprule
        $W\%$ & \multicolumn{3}{c|}{Load$=100\%$} & \multicolumn{3}{c|}{Load$=105\%$} & \multicolumn{3}{c}{Load$=110\%$} \\
    \cmidrule(lr){2-10}
   UF & \#CC-V & \#I & time & \#CC-V & \#I & time & \#CC-V & \#I & time \\
    \midrule
    20 & 7 & 3 & 364.46 & 8 & 4 & 3131.63 & 8 & 2 & 388.73 \\
    30 & 7 & 2 & 461.95 & 8 & 2 & 741.27 & 7 & 2 & 373.30 \\
    40 & 7 & 2 & 536.97 & 7 & 2 & 557.93 & 7 & 2 & 460.86 \\
    50 & 4 & 2 & 933.50 & 3 & 2 & 390.23 & 4 & 2 & 363.07 \\
    \bottomrule
    \end{tabularx}
    % \end{tabular}
\end{table}

\subsection{Out-of-sample tests} \label{subsec:outofsampletests}
Next, we study the performance of the dispatch computed by the SCCUC  when there are errors in the assumption that wind power follows a normal distribution by performing an out-of-sample test. The formulation and algorithm for the SCCUC assume that the wind power generated is normally distributed \cite{Dvorkin2016}, but studies in the literature suggests wind energy distribution does not always follow a normal distribution \cite{Seguro2000}. Here, we test the generation dispatch computed by the SCCUC using forecast of mean wind energy production and variance in wind energy produced by each wind farm (assumption that the wind is normally distributed) against samples from other distributions for the wind power generation. 
We use samples from the following probability distributions, all with fatter tails than the normal distribution: (i) Laplace, (ii) Logistic, and (iii) Weibull with 2 different shape parameters as in \cite{Karki2006}. For the Laplace and Logistic distributions, we match the mean and standard deviation estimated from the forecast data by assuming a normal distribution. For the Weibull distribution, we consider shape parameters $k = 1.2, 2$ and choose the scale parameter to match the standard deviations; the samples are then translated to match the means. The dispatch from the SCCUC is tested against $1000$ realizations for each distribution and the maximum empirical violation probabilities, evaluated for each constraint separately, are tabulated in Table \ref{tab:EmpViol}. 

We observe that the dispatch obtained using the SCCUC solution performs the worst for the realizations from the Weibull distributions. This is not surprising considering the fact that the Weibull distribution is a bad approximation of a normal distribution. For all the other cases, the maximum probability of violations are well within $1.8\%$ and $20\%$ for the generators and lines, respectively. We note that these violations do not include the ones that occur during single line outages and that similar trends were observed for the line limits during line outages. 

\begin{table}[htbp]
    \caption{Maximum violation probability, \kaarthik{in percentage}, for out-of-sample tests with $\varepsilon_i = 1\%$ and $\varepsilon_{\ell}=10\%$ for the  dispatch given by the SCCUC.}
     \label{tab:EmpViol}
    \centering
    \footnotesize
    \begin{tabular}{l|cc}%cc}
        \toprule
        Distribution & Generators & Lines \\
        \midrule
         Laplace  & 0.0052 & 0.0910 \\
         Logistic & 0.0015 & 0.0732  \\
         Weibull, $k=1.2$ & 0.0212 & 0.2531 \\
         Weibull, $k=2$ & 0.0181 & 0.1999 \\
         \bottomrule
    \end{tabular}    
\end{table}

We remark that tighter or looser violation probability values $\varepsilon_i$, $\varepsilon_{\ell}$ with the normal distribution assumption, can be chosen if prior knowledge of the probability distribution of wind is obtained. For instance, if one is certain that the probability distribution of wind is given by a Weibull distribution, the SCCUC can still be utilized by choosing very conservative values for $\varepsilon_i$ and $\varepsilon_{\ell}$. 

\subsection{Comparison of the SCCUC to its deterministic counterpart}
From an application perspective, the comparison between SCCUC and its deterministic counterpart is valuable for power system operators. The deterministic counterpart assumes $\boldsymbol \omega_b(t)=0$. Both the deterministic and chance constrained unit commitment with N-1 security constraints are solved for a case where the forecast wind power production accounts for $20\%$ of the total load. Since the deterministic unit commitment with N-1 security constraints assumes that there are no wind fluctuations, it will not require any generation reserves to handle the wind power fluctuations.  In order to make a fair comparison, we assume that the system operator maintains a minimum generation reserve requirement (nominal reserve rule) to handle load and wind power fluctuations. \kaarthik{This reserve requirements is defined such that it corresponds to the reserve capacities required to balance wind power fluctuations in the chance constrained formulation \cite{Roald2016}, i.e., 
%equal to $0.5\%$ of the load and add this constraint to the optimization problem:
\begin{flalign*}
&\sum_{i\in \mathcal G} r_i^-(t) \geqslant \phi^{-1}(1-\epsilon_i) \sqrt{e^T \Sigma(t) e} \quad \forall t\in \mathcal T, \\
&\sum_{i\in \mathcal G} r_i^+(t) \geqslant \phi^{-1}(1-\epsilon_i) \sqrt{e^T \Sigma(t) e} \quad \forall t\in \mathcal T.
%&\sum_{i\in \mathcal G} r_i^+(t) \geqslant 0.005 \cdot \sum_{b \in \mathcal B} d_b(t) \quad \forall t\in \mathcal T \\
%&\sum_{i\in \mathcal G} r_i^-(t) \geqslant 0.005 \cdot \sum_{b \in \mathcal B} d_b(t) \quad \forall t\in \mathcal T.
\end{flalign*}
The above constraints ensure that the amount of reserves allotted to handle wind fluctuations in the deterministic problem is essentially equal to the reserves allotted using a chance-constrained formulation. }
% where $R^+(t)=R^-(t)=\Phi^{-1}(1-\epsilon_i)\sigma_\Omega$.}
Furthermore, to obtain a more interesting case, the transmission capacities of the IEEE RTS-96 system were decreased to $90\%$ of their original base case value in \cite{Pandzic2015}. 

\subsubsection{Cost comparison}
The resulting deterministic unit commitment problem with $N-1$ contingencies is a mixed-integer linear program that is solved using CPLEX 12.7. The time taken by CPLEX to produce a feasible solution with a $1\%$ optimality tolerance for the deterministic problem is $253.53$ seconds. The modified Benders decomposition algorithm solves the same instance \kaarthik{in} $525.31$ seconds.   
The total cost of the unit commitment and the different cost components are shown in Table \ref{tab:costs}. We observe that the total cost of the SCCUC solution is almost equal to the cost of its deterministic counterpart. \kaarthik{The main takeaway from Table \ref{tab:costs} is that, the chance constrained unit commitment with N-1 constraints can accommodate fluctuations in wind while protecting the system against contingencies without incurring too much in operational costs. We remark that the reserves from $r_i^+$ and $r_i^-$ are exactly the same in both the deterministic and the chance-constrained problems because the deterministic counterpart was enforced to allocate exactly the same amount of reserves as that of the chance-constrained problem.  But, in practice, a nominal reserve rule of $3\%$ to $5\%$ of the total load in the system is allocated for reserves to account for the wind fluctuations. This may be more or less than what is required to handle the wind fluctuations depending on the forecast values. Furthermore, the exact generators and their individual contribution to these reserve are not optimized in the deterministic formulation. Hence, the chance-constrained optimization problem proposed in this paper provides a more economical and accurate solution to handle wind penetration into the transmission system while keeping the system secure under N-1 contingencies.}

\begin{table}[htbp]
    \caption{Deterministic vs. Chance-constrained (SCCUC) comparison}
    \label{tab:costs}
    \centering
    \footnotesize
    \begin{tabular}{r|ll}
         \toprule
         & Deterministic & Chance-constrained  \\
         \midrule
         Total operational cost (\$) & \kaarthik{$2.026\times10^6$} & $2.029\times10^6$\\
         No load cost (\$) & $3.141\times 10^5$ & $3.015\times 10^5$ \\
         Start-up cost (\$) & $14918.5$ & $27077.4$ \\
         Production cost (\$) & $1.676\times 10^6$ & $1.679\times 10^6$ \\
         \linecomment{Reserve capacity $r_i^{out}$ cost} (\$) & $19115.2$ & $19141.89$ \\
         \linecomment{Reserve capacity $r_i^+$, $r_i^-$ cost} (\$) & \kaarthik{$2966.60$} & $2966.60$ \\
         Optimality gap (\%) & $0.57$ & $0.38$ \\
         \linecomment{Reserve capacity $r_i^+$, $r_i^-$ (MW)}  & \kaarthik{$1483.33$} & $1483.33$ \\
         \bottomrule
    \end{tabular}
\end{table}

\subsubsection{Line and generation limit violations}
\begin{figure}[htbp]
    \centering
    \includegraphics[scale=0.3]{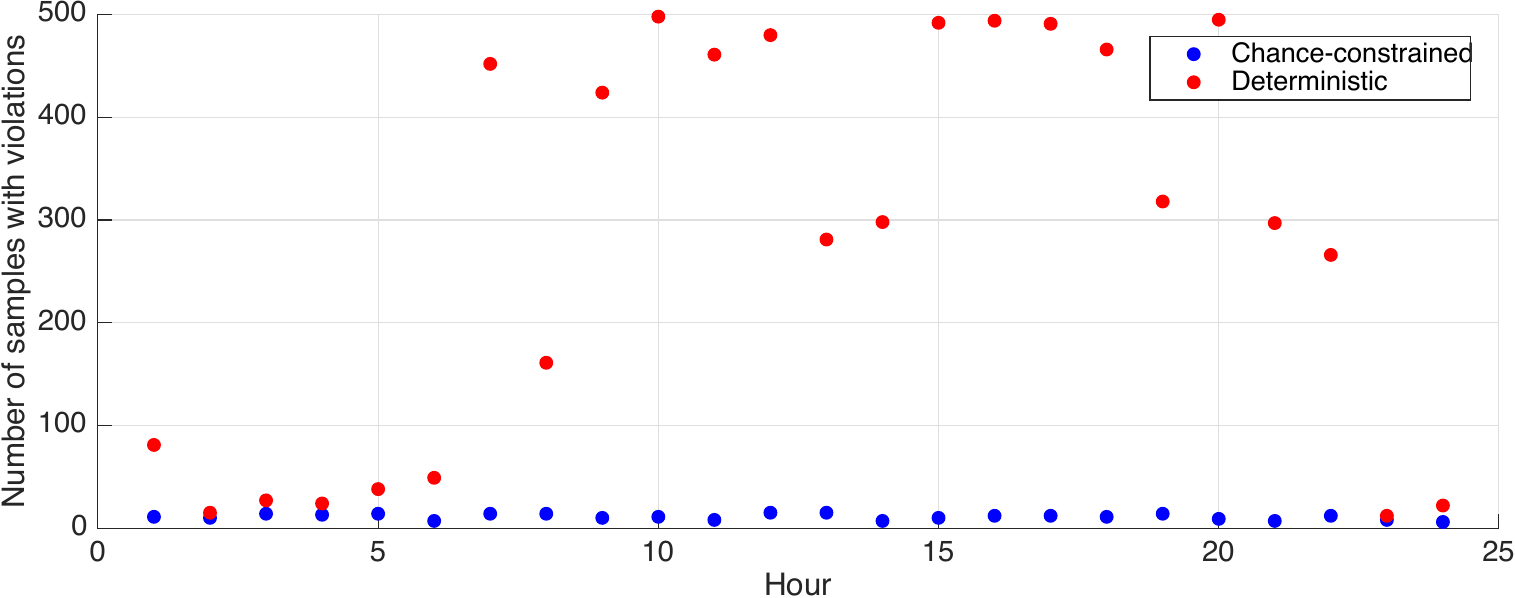}
    \caption{The number of wind realizations from the logistic distribution (out of 1000) that lead to one or more constraint violations for the deterministic and the chance-constrained (SCCUC) solutions, respectively.}
    \label{fig:SampleViol}
\end{figure}

For a system operator, it is also important to reduce the frequency that \emph{any} line or generator limit is violated during a 24-hour period. Fig. \ref{fig:SampleViol} compares the number of wind sample realizations out of 1000 that lead to at least one constraint violation during each hour of the day.  From  Fig. \ref{fig:SampleViol}, the dispatch obtained from the deterministic problem indicates the empirical probability that at least one constraint is violated during each hour is between $3\%$ and $50\%$. The solution obtained from the SCCUC has a significantly smaller empirical probability.  

\subsection{Influence of wind penetration}
To assess the influence of wind power penetration we compare the cost of unit commitment and amount of reserves for varying levels of wind energy penetration. Table \ref{tab:windlevels} shows the variation of the number of committed units, \linecomment{the reserve capacity procured to balance wind power fluctuations} and the total SCCUC cost for different levels of wind penetration. We observe that the increase in the wind penetration levels decrease the overall SCCUC cost. The \linecomment{reserve capacity} remains more or less constant because the variance in the wind deviations is a constant over all instances for this particular study. The decrease in the number of units committed can also be explained by the fact that power production from the conventional generators in the system decreases with increasing levels of wind penetration.

\begin{table}[htp]
\caption{Number of committed units, reserve capacities $\sum (r_i^+ + r_i^-)$ (MW), and SCCUC cost (\$) for varying levels of wind penetration}
\footnotesize
\centering
\setlength\tabcolsep{2pt}
\begin{tabularx}{\columnwidth}{ccccc}
% \begin{tabular}{ccccc}
\toprule
 Wind  & Committed & Wind Fluctuation & Total SCCUC & Computation\\
 penetration [\%] & Units & Reserves [MW] & Cost [$\times 10^6$\$] & time in sec.\\
 \midrule
 50 & 699   & 296.60 & 1.5656 & 933.50\\
 40 & 717   & 296.63 & 1.6914 & 536.97 \\
 30 & 826   & 296.66 & 2.0015 & 461.95\\
 20 & 923   & 296.60 & 2.3509 & 364.46\\
 \bottomrule
% \end{tabular}
\end{tabularx}
\label{tab:windlevels}
\end{table}

\section{Conclusion}

In this paper, we have presented a MISOCP formulation for the SCCUC problem in the presence of wind fluctuations. To the best of our knowledge, this is the first UC formulation in the literature that includes N-1 security constraints on lines and generators, wind fluctuations, and reserve capacities to balance both wind fluctuations and outages. A modified Benders decomposition algorithm is developed to solve the problem. The effectiveness of the approach and its advantages over its deterministic counterpart was demonstrated through extensive computational experiments on a variation of the IEEE RTS-96 system. 
The results indicate that the proposed formulation is effective and can result in better technical performance including fewer violations of transmission line limits and generator limits during normal operations and during single line or generator outages when compared to its deterministic counterpart. The algorithm was also shown to scale to systems with up to 300 nodes. We argue that the results on this network size, combined with N-1 constraints, demonstrate that our approach is most scalable algorithm for UC with uncertain wind power to date. 

Future work includes generalizations to account for errors in estimating the parameters of the probability distributions for the wind fluctuations and extension of the SCCUC formulation to consider more realistic (AC) power flow models. The work on this problem has also provided motivation to improve the quality of conic optimization solvers and their ability to produce reliable duals. It will also be interesting to understand how the approaches we developed here for the SCCUC could be extended and generalized to other two-stage MISOCP problems.

\section*{Acknowledgements} This work was supported by the Advanced Grid Modeling Program of the Office of Electricity within the U.S Department of Energy and the Center for Nonlinear Studies at Los Alamos National Laboratory. We also thank Miles Lubin for his input and contributions on early iterations of these research efforts.

\bibliographystyle{unsrt}
\bibliography{references.bib}

\begin{IEEEbiography}[
{\includegraphics[width=1in,height=1.25in,clip,keepaspectratio]{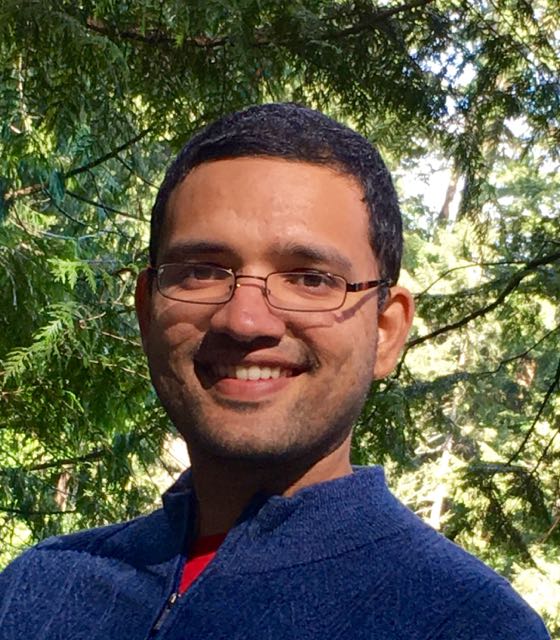}}
]{Kaarthik Sundar}
received the Ph.D. degree in mechanical engineering from Texas A\&M University, College Station, TX, USA, in 2016. He is currently a Research Scientist in the Information Systems and Modeling Division at Los Alamos National Laboratory, Los Alamos, NM, USA. His research interests include problems pertaining to vehicle routing, path planning, and control for unmanned/autonomous systems; numerical optimal control, estimation, and large-scale optimization problems in power and gas networks; combinatorial optimization; and global optimization for mixed-integer nonlinear programs.
\end{IEEEbiography}

\begin{IEEEbiography}[
{\includegraphics[width=1in,height=1.25in,clip,keepaspectratio]{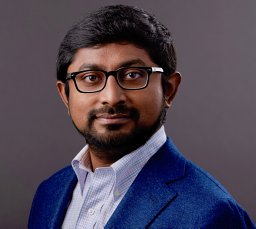}}
]{Harsha Nagarajan}
received the Ph.D. degree in mechanical engineering from Texas A\&M University, College Station, in 2014, in the area of Controls and Optimization. He is currently a Scientist with the Applied Mathematics and Plasma Physics Division, Los Alamos National Laboratory, Los Alamos, NM, USA. His research interests include solving a myriad of problems in the fields of control theory and (discrete/nonconvex) optimization applied to energy infrastructure systems and UAV routing.
\end{IEEEbiography}

\begin{IEEEbiography}[
{\includegraphics[width=1in,height=1.25in,clip,keepaspectratio]{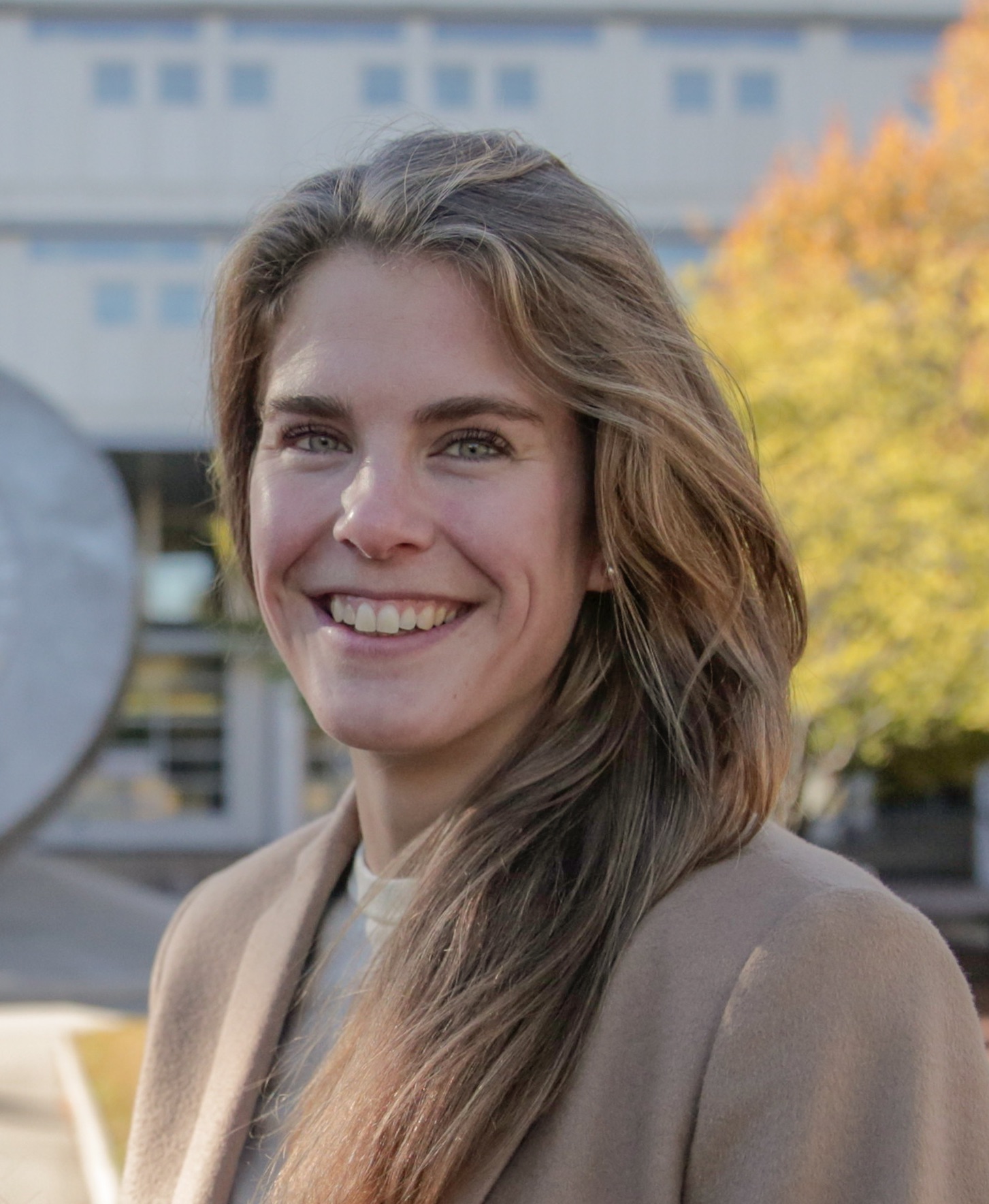}}]{Line Roald}
received her Ph.D. degree from ETH Zurich, Zurich, Switzerland in 2016. She is currently an Assistant Professor and a Grainger Institute Fellow in the Department of Electrical and Computer Engineering in University of Wisconsin-Madison. Her research interests focus on modeling and optimization of energy systems, and particularly on managing uncertainty and risk arising from renewable energy variability and component failures.
\end{IEEEbiography}

\begin{IEEEbiography}[
{\includegraphics[width=1in,height=1.25in,clip,keepaspectratio]{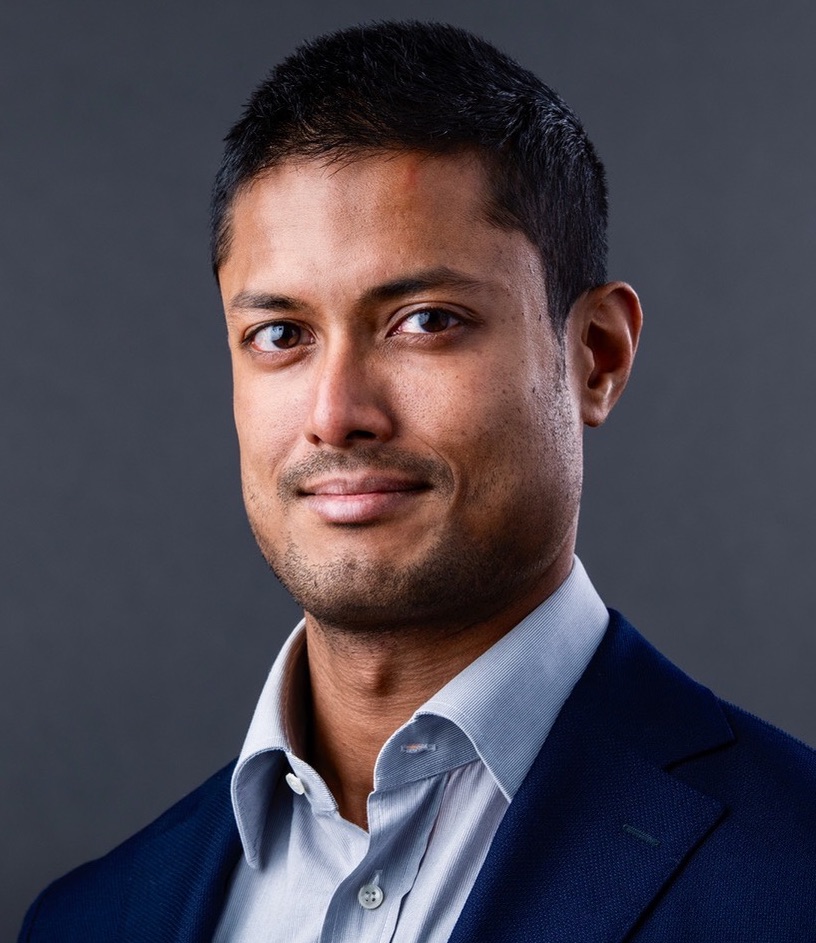}}
]{Sidhant Misra}
Sidhant Misra received the Ph.D. degree in electrical engineering and computer science from the Massachusetts Institute of Technology, Cambridge, MA, USA, in 2014. He is currently a Research Scientist in the theory division and the Advanced Network Science Initiative at the Los Alamos National Laboratory, Los Alamos, NM, USA. His research interests lie at the intersection of machine learning and optimization with applications in network structure identification, and planning and operations of energy networks
\end{IEEEbiography}

\begin{IEEEbiography}[
{\includegraphics[width=1in,height=1.25in,clip,keepaspectratio]{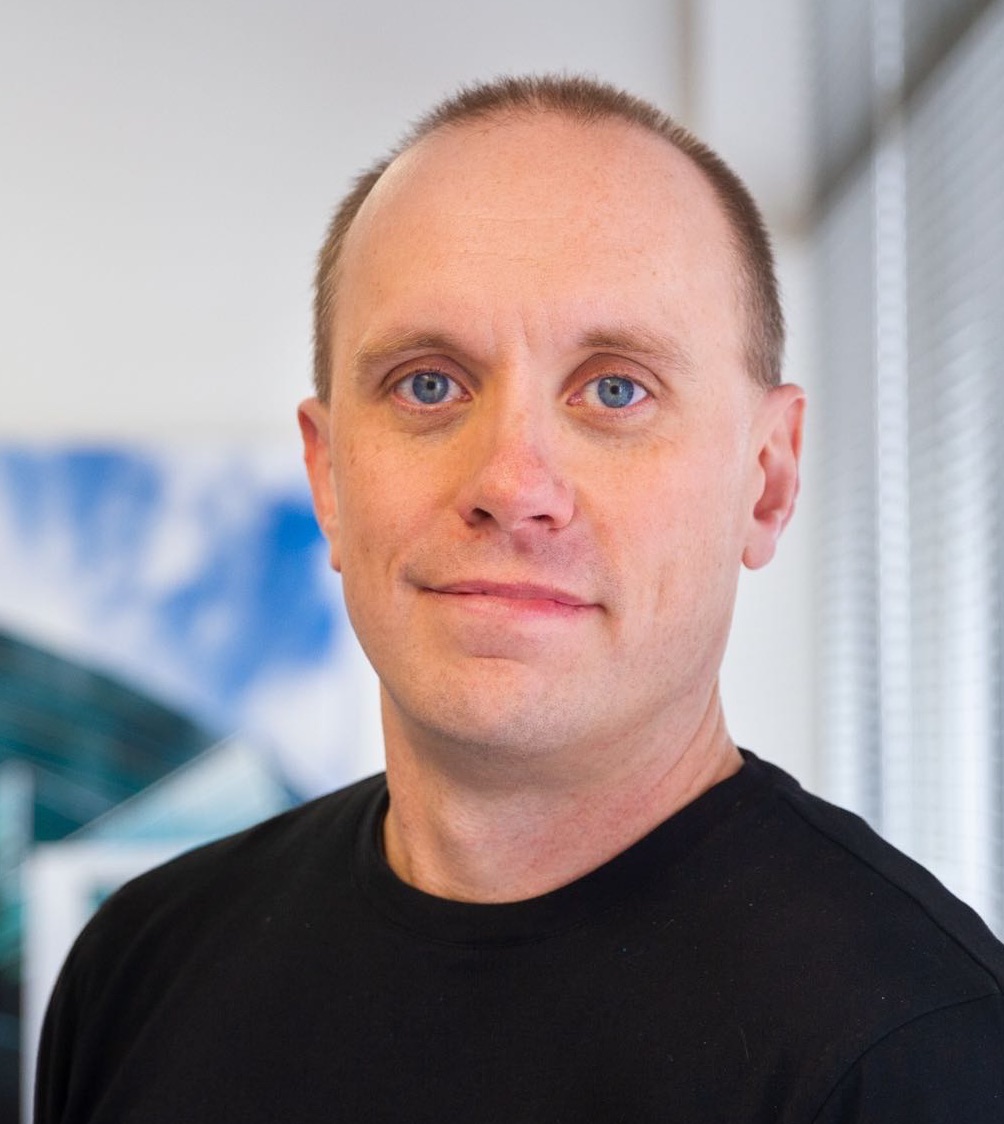}}
]{Russell Bent}
Russell Bent received his Ph.D. degree in computer science from Brown University
in 2005. Since then he has been a scientist at Los Alamos National Laboratory, NM,
USA. He is currently in the Applied Mathematics and Plasma Physics Group (T-5),
where he leads LANL's inter-organizational Advanced Network Science Initiative. 
His research interests include modeling inter-dependencies between energy infrastructure systems, managing
disasters that impact critical infrastructure, modeling smart grid technologies, and
developing methods for mixed-integer, non-linear optimization.
\end{IEEEbiography}

\begin{IEEEbiography}[
{\includegraphics[width=1in,height=1.25in,clip,keepaspectratio]{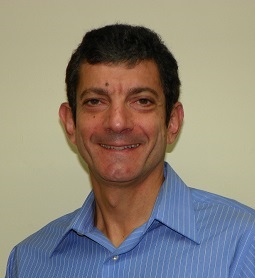}}
]{Daniel Bienstock}
is Liu Family Professor of Operations Research at Columbia University.  His research focuses on theory and computational aspects of optimization, and in problems related to electrical transmission systems.  He is an Informs Fellow.
\end{IEEEbiography}

\end{document}